\documentclass[%
 aip,
 amsmath,amssymb,
 reprint,%
]{revtex4-1}
\usepackage[utf8]{inputenc}
\usepackage[T1]{fontenc}
\usepackage{mathptmx}
\usepackage{amsmath}
\usepackage{lipsum}
\usepackage{esint}
\usepackage{graphicx,subfigure}
\usepackage{siunitx}
\usepackage{csquotes}
\usepackage{textcomp}
\usepackage{dcolumn}
\usepackage{bm}
\usepackage{array} 
\usepackage{xcolor}
\usepackage[colorlinks=true,urlcolor=cyan,citecolor=cyan,linkcolor=cyan,bookmarks=true]{hyperref}
\usepackage{rotating}
\usepackage{array}
\usepackage{booktabs} 
\usepackage{multirow} 
\usepackage{lineno}

\usepackage{placeins}

\DeclareMathAlphabet{\mathsfit}{T1}{\sfdefault}{\mddefault}{\updefault}
\SetMathAlphabet{\mathsfit}{bold}{T1}{\sfdefault}{\bfdefault}{\updefault}
\renewcommand{\vec}[1]{\mathbf{{#1}}}
\renewcommand{\cite}[1]{\citep{#1}}

\newlength{\FigureHeight}
\newlength{\FigureHeightHalf}

\begin{document}

\preprint{Physics of Fluids}

\title{A Computational Fluid Dynamics MacroModel for the Design of Bed Adsorbers}

\author{Mohamad Najib Nadamani}
\affiliation{CNRS, CORIA, UMR 6614, University of Rouen Normandy, Rouen, F-76000, France}

\author{Mostafa Safdari Shadloo}
\affiliation{INSA Rouen Normandie, Univ Rouen Normandie, CNRS, Normandie Univ, CORIA UMR 6614
, F-76000, France}
\affiliation{Institut Universitaire de France, Paris, F-75231, France}

\author{Talib Dbouk}
\email{talib.dbouk@coria.fr, Corresponding Author}
\affiliation{CNRS, CORIA, UMR 6614, University of Rouen Normandy, Rouen, F-76000, France}


\begin{abstract}
A new three-dimensional (3D) multiphase computational fluid dynamics (CFD) model for adsorption physics in packed beds of spherical beads is developed and validated. The model is constituted at a macroscopic scale that integrates new volumetric source terms in the multi-species gas transport and energy conservation equations. These new terms, for the first time, take into account the impact of pores adsorption occupation rate (PAOR), or gas loading.
Transient 3D simulations are performed at an atmospheric pressure of about 1.02 bar for different CO\textsubscript{2}-He gas mixture feed-in compositions (100\%, 50\%, and 15\% CO\textsubscript{2}). The 3D model validation is conducted through quantitative comparisons with experimental data from the literature for CO\textsubscript{2} adsorption on porous Zeolite-13X beads in a cylindrical fixed-bed. Results demonstrate the new model’s ability to accurately predict the breakthrough curves and the thermal front propagation inside the bed. Finally, the new CFD model is applied to investigate CO\textsubscript{2} capture in a new 3D design of fixed-bed adsorbers of equivalent adsorbent material volume. The new design outperformed the reference cylindrical design thanks to its new geometry with higher surface area. This allows to shorten the adsorption periods in pressure and temperature swing adsorption processes and thus increase the overall gas separation process productivity.
\end{abstract}


\maketitle


\begin{table}[h!]
\caption{List of nomenclature and description.}
\begin{tabular}{ll}
\hline
\textbf{Nomenclature} & Description \\ \hline
$b_i$ & Langmuir constant for component $i$  [m$^3$/mol] \\
$C$ & Concentration of adsorbate in the gas phase [mol/m$^3$] \\
$c_{i}$ & Molar concentration of component $i$  [mol/m$^3$] \\
$c_{p}$ & Specific heat capacity of the gas phase  [J/(kg·K)] \\
$c_{p_{eff}}$ & Effective specific heat capacity of the bed [J/(kg·K)] \\
$c_{p_{p}}$ & Specific heat capacity of the solid particles  [J/(kg·K)] \\
$d_i$ & Langmuir afnity constant for component $i$ [m$^3$/mol] \\
$d_p$ & Particle diameter [m] \\
$D_L$ & Axial dispersion coefficient [m$^2$/s] \\
$D_m$ & Binary molecular diffusion coefficient [m$^2$/s] \\
$Gr$ & Grashof number [-] \\
$h$ & Convective heat transfer coefficient [W/(m$^2$·K)] \\
$i$ & Species number or gas component \\
$k_i$ & LDF mass transfer coefficient [1/s]\\
$M_i$ & Molar mass of species $i$ [kg/mol]\\
$p$ & Pressure [bar] \\
$Pr$ & Prandtl number [-]\\
$q$ & Adsorbed quantity per unit mass [mol/kg] \\
$r$ & Cylindrical bed's radius [m] \\
$R$ & Universal gas constant [J/(mol·K)] \\
$S_T$ & Energy conservation equation source term [J/(m$^3$·s)] \\
$S_Y$ & Species-transport equation source term [kg/(m$^3$·s)] \\
$t$ & time [s] \\
$T$ & Temperature [K] \\
$T(z,t)$ & Wall Temperature (z,t)-dependent [K] \\
$T_{ref}$ & Ambient Temperature [K] \\
$u$ & Velocity [m/s] \\
$(x,y,z)$ & Cartesian local coordinates [m] \\
$Y_i$ & Mass fraction of species $i$ [-] \\
\hline
\end{tabular}\label{tableI}
\end{table}

\newpage

\begin{table}[h!]
\caption{List of Greek letters symbols and description.}
\begin{tabular}{ll}
\hline
\textbf{Greek Letters Symbol} & Description \\ \hline
$\kappa$ & Gas thermal conductivity [W/(m·K)] \\
$\kappa_{eff}$ & Effective thermal conductivity [W/(m·K)] \\
$\kappa_p$ & Solid thermal conductivity [W/(m·K)] \\
$\rho_g$ & Gas phase density [kg/m$^3$] \\
$\rho_{eff}$ & Effective density [kg/m$^3$] \\
$\rho_p$ & Solid adsorbent density [kg/m$^3$] \\
$\varepsilon_p$ & Particle porosity [-] \\
$\varepsilon_b$ & Bed porosity [-] \\
$\varepsilon_t$ & Total porosity [-] \\
$\Gamma_Y$ & PAOR term [-] in eqn. \ref{equ:DeltaYspeciesConservation}\\
$\Gamma_T$ & PAOR term [-] in eqn. \ref{equ:DeltaTenergyConservation}\\
$\mu$ & Dynamic viscosity [Pa·s] \\
$\Delta H$ & Isosteric heat of adsorption [J/mol] \\
$\Sigma_{v,i}$ & Constants, eqn.\ref{DmCO2} ($\Sigma_{v,CO2}=26.9$, $\Sigma_{v,He}=2.88$) \\
\hline
\end{tabular}\label{tableII}
\end{table}

\begin{table}[h!]
\caption{List of abbreviations and description.}
\begin{tabular}{ll}
\hline
\textbf{Abbreviations} & Description \\ \hline
1D & One-dimensional\\
2D & Two-dimensional\\
3D & Three-dimensional\\
CCS & Carbon capture and storage\\
CFD & Computational Fluid Dynamics\\
CFL & Courant–Friedrichs–Lewy \\
CO$_2$ & Carbon Dioxide\\
DSL & Dual-Site Langmuir\\
FVM & Finite Volume Method\\
GCI & Grid Convergence Index\\
He & Helium\\
LDF & Linear Driving Force\\
NIST & National Institute of Standards and Technology\\
PAOR & Pores Adsorption Occupation Rate\\
PSA & Pressure Swing Adsorption\\
TSA & Temperature Swing Adsorption\\
\hline
\end{tabular}\label{tableIII}
\end{table}

\section{Introduction}\label{sec:intro}

Carbon dioxide (CO\textsubscript{2}) concentrations have been increasing in atmosphere in recent decades where at least two-third of greenhouse effect gas emissions are caused by different human activities.
Moreover, natural gas and coal based fired power plants account for huge amounts of energy production that produces billion metric tons of CO\textsubscript{2} worldwide, contributing to the overall global warming. Governmental strategies at worldwide levels, e.g. as Paris climate change agreements, have been developed in attempts such that nations must urgently decrease CO\textsubscript{2} emissions. Gas separation such as carbon capture and storage technologies (CCS) are promising solutions to reduce CO\textsubscript{2} and greenhouse effect gas emissions \cite{Songolzadeh2014}.

Adsorption based technologies for CO\textsubscript{2} capture have been an important topic for research and developments in the last decade. For example, Song et al. 2025 \cite{Song2025} employed granulation technology to improve the performance of Alkaline metal salt-promoted MgO sorbents as effective materials for CO\textsubscript{2} capture. Liu et al. 2024 \cite{LIU:2024} investigated CO\textsubscript{2} capture technology using pressure swing adsorption (PSA) for the petrochemical industry. Das et al. 2023 \cite{LIU:2024} presented a comprehensive review adsorption based CO\textsubscript{2} capture technologies. Riboldi et al. 2017 \cite{RIBOLDI20172390} and Siqueira et al. 2017 \cite{SIQUEIRA20172182} presented state of the art overviews on PSA as an innovative CO\textsubscript{2} capture technology. White at al. 2016 \cite{White2016} development a PSA cycle to capture CO\textsubscript{2} from flue gas employing a 4-bed PSA apparatus. \textsubscript{}

As can be seen, PSA technology, is thus emerging as a promising solution for post‐combustion CO\textsubscript{2} capture. This is thanks to its relatively low energy requirements and operational modularity options. In a PSA cycle, a fluid mixture is passed through a packed bed (usually fixed-bed adsorber or reactor) containing porous solid sorbent: CO\textsubscript{2} preferentially adsorbs onto the sorbent surfaces (nano-pores nanostructure) under high pressure (adsorption step), and the bed is then depressurized to desorb (regenerate) the sorbent and recover the captured CO\textsubscript{2} \cite{Yang1984, Ruthven1994}, that can be transported or stocked underground. 

Fixed-bed adsorption systems, in particular, are highly valued for their operational simplicity, robustness, and adaptability to a wide range of flow and concentration conditions. However, the effective design and optimization of such systems hinge on a deep understanding of the coupled phenomena governing adsorption, which include mass transfer, fluid dynamics, heat effects, and sorbent kinetics. 

Numerical modeling and simulations have been rapidly progressing in the last years. For example, Artificial Intelligence is a promising window for {future chemical engineering technologies \cite{Aoming2025, kashefi2021point}}. The development of fractal dynamics models for adsorption is also very promising \cite{Zhang2025,Zhou2025}. {Adsorption modeling has been also developing for solutes adsorption. For example, Meng et al. (2026) \cite{meng2026reactive} investigated through modeling and simulations the reactive dispersion dynamics in packed tube flow with wall adsorption and desorption.} In the context of physics-based numerical modeling and simulation, computational fluid dynamics (CFD) has emerged as a powerful tool to simulate the transient and spatially varying nature of these processes. In the last decade, CFD-based models have increasingly been developed in attempts to better predict the local dynamic behavior of adsorption kinematics and temperature front transport within fixed packed beds. Verbruggen et al. 2016 \cite{verbruggen2016cfd} demonstrated the capability of CFD to simulate gas-phase transport and reaction phenomena in catalytic and adsorptive systems. Kasai et al. 2023 \cite{kasai2023cfd} and Gautier et al. 2018 \cite{gautier2018pressure,GAUTIER2018314} developed CFD models for PSA to predict the adsorption rate in packed beds of granular activated carbon. Ramos et al. 2024 \cite{ramos2024cfd} developed a CFD model is validated with experimental measurements of adsorption of CO\textsubscript{2} in-gas mixtures employing packed beds of Zeolite-13X. 

The strength of CFD lies in its ability to resolve local gradients in temperature, pressure, and concentration, information that is difficult to obtain experimentally and often neglected in lumped-parameter or one-dimensional (1D) adsorption models \cite{HWANG1995,da1999general}. The research progress in this field underlined that three-dimensional (3D) simulations are essential when realistic prediction of local gas dynamics and heat transfer is needed, particularly in beds with non-conventional shapes \cite{ben2017multicomponent} or multi-dimensional heat loss paths. In adsorption-based CO\textsubscript{2} capture systems, understanding such local effects is vital for enhancing breakthrough curves and reducing maximum temperature local spikes, for enhanced bed efficiency and increased productivity.

{In this work, we develop and implement a new robust 3D CFD model using the \href{https://www.openfoam.com}{\textbf{OpenFOAM}} platform, for gas adsorption in fixed-bed adsorber with an application to CO\textsubscript{2} capture. The solver incorporates volume-averaged conservation equations for mass, momentum, and energy, including new source terms that take into account the impact of pores adsorption occupation rate (PAOR), or gas loading, in the CFD modeling.} Zeolite 13X is used as the adsorbent material \cite{son2018,chen2017co2}, given its well-documented selectivity and capacity for CO\textsubscript{2} \cite{Hauchhum_Mahanta_2014}. The new model is validated against experimental data from the literature by Wilkins and Rajendran \cite{wilkins2019measurement} and Ramos et al.~\cite{ramos2024cfd}, for three different CO\textsubscript{2} feed-in concentrations (100\%, 50\%, and 15\%) at atmospheric pressure of $1.02$ bar. 

{Unlike most prior studies that assume cylindrical bed geometries, we further extend the 3D CFD solver's capabilities by investigating a more complex multi-tube bed design, which maintains the same total adsorbent volume. This highlights from one hand the importance of 3D CFD models in capturing realistic adsorption and thermal front dynamics, and on the other hand their potential to design next generations of fix-bed adsorbers.}

{The present 3D CFD model is limited to gas mixture flows in adsorbing porous media under the assumption that all the gas phases are in local thermal equilibrium with the solid phase. In other words, the present model assumes one effective temperature equation.}

\section{Mathematical Modeling and Governing Equations}
\label{sec:math_model}
\subsection{Model description}

In this section, we present the developed comprehensive 3D CFD model to simulate the physical adsorption process within a fixed-bed adsorber (a reference design from the literature, and a new 3D proposed enhanced design). This is based on \href{https://www.openfoam.com}{\textbf{OpenFOAM}}-v2306 open source platform. The geometry of the reference design of the fixed-bed adsorber is constructed to replicate the bed in the experimental setup presented by Wilkins and Rajendran \cite{wilkins2019measurement}. It comprises a cylindrical fixed-bed made of an internal diameter of 2.82 cm and a total length of 6.4 cm. 

In CFD modeling of adsorption physics, the coupled conservation equations of mass, momentum, and energy must be solved simultaneously. These conservation equations form the background basis for describing the key transport phenomena within fixed-bed adsorption columns. To accurately model the fluid flow, heat transfer, and mass transport phenomena within the packed bed, the porous medium can be modeled at different scales, $i.e.$ at the scale of pellets/beads or at a continuum macroscopic scale. In the latter approach, the porous structure of the bed is implicitly modeled using averaged quantities such as porosity and particle diameter. This allows for more computationally affordable computations without resolving the physics at the scale of each individual particle (pellet or bead). The modeling framework in the present work adopts this macroscopic scale modeling of the adsorbing fixed-bed but with additional improvements to take into account the gas loading impact of PAOR in the CFD modeling as we will explain in section \ref{SY_ST}.

{In this work, some assumptions are adopted while maintaining numerical accuracy:}
\begin{itemize}
    \item { A laminar flow regime at the scale of the adsorbing particles such that $Re_p \approx 1.7$.}
    \item  {A local thermal equilibrium at the solid-gas interface.}
    \item {An Ideal gas law of thermophysical properties}.    
    \item {A bi-molecular diffusion model.}
\end{itemize}

The proposed framework solves for the multispecies gas Navier–Stokes (momentum) equations, continuity (mass) equation, species transport and the energy equation, all under transient and compressible multispecies gaseous flow conditions. To account for adsorption, additional sub-models are integrated as will be explained. These include the evaluation of both saturated and instantaneous adsorption capacities, which influence the local mass and energy balances through species uptake and heat effects. 
Mass transfer mechanisms and heat exchange between the fluid and solid phases are modeled by incorporating new two source terms into the governing mass and energy conservation equations, while pressure drop is accounted for by adding momentum source terms that represent viscous and inertial losses in the porous medium. Furthermore, adsorption kinetics and thermal effects due to exothermic adsorption reactions are incorporated to enhance the physical realism of the simulation. Together, these components are coupled to yield a fully integrated CFD–adsorption model. This enables us to accurately predict the spatio-temporal evolution of species concentrations, temperature profiles, and pressure fields during the adsorption period.

The governing equations in the present 3D CFD model consist of the conservation laws for mass, momentum, energy, and gaseous species transport, expressed as the following:

\subsubsection{Momentum conservation} 
    The momentum conservation equation is defined as:
        \begin{equation}
     \frac{\partial(\rho_g \mathbf{u})}{\partial t} + \nabla \cdot (\rho_g \mathbf{u} \mathbf{u}) - \nabla \cdot \mathbf{\tau} =  - \overrightarrow{\nabla p}
    \end{equation}

$\mathbf{\tau}$ is the stress tensor such that:
    \begin{equation}
    \mathbf{\tau} = - {2 \over 3} \mu_{g} \left(\nabla \cdot \mathbf{u} \right) \mathbf{I} +  \mu_{g} \left(\nabla  \mathbf{u} + {(\nabla  \mathbf{u})}^T  \right)
    \end{equation}

$\mu_{g}$ is the effective dynamic viscosity of the gas mixture and $\mathbf{I}$ is the identity tensor. Note that all symbols and abbreviations are summarised in Tables \ref{tableI}, \ref{tableII} and \ref{tableIII}.

    The pressure gradient across a packed bed of porous medium of spherical particles, can be modeled by the Ergun's equation \cite{ergun1952} as the following:
    \begin{equation} \label{equ:ergun}
        \overrightarrow{\nabla p} = -\frac{150 \mu_{g} (1 - \varepsilon_b)^2}{\varepsilon_b^3 d_p^2} \mathbf{u} - \frac{1.75 (1 - \varepsilon_b) \rho_g}{\varepsilon_b^3 d_p} |\mathbf{u}| \mathbf{u}
    \end{equation}
    where $t$ is the time, $p$ is the pressure scalar field, $\mathbf{u}$ is the velocity vector field, $d_p$ is the particle diameter, $\rho_g$ its density and $\varepsilon_b$ is the bulk bed porosity.

\subsubsection{Mass conservation}
The adsorption process entails a transfer of mass from the gas phase to the solid phase, i.e. adsorbent. The mass conservation is governed by:

\begin{equation}\label{equ:massConservation}
      \frac{\partial \rho_g}{\partial t}
      + \nabla \cdot (\rho_g \mathbf{u})
      =
      S_Y
\end{equation}

Where $S_Y$ is a sink term that accounts for the reduction in the gas-phase mass due to adsorption.

\subsubsection{Species transport conservation}
The conservation of each chemical species within the gas phase is governed by a species transport equation that accounts for advection and dispersion of each gas component inside the bed column. 

\begin{equation} \label{equ:speciesConservation}
      \frac{\partial \rho_g Y_i}{\partial t}
      + \nabla \cdot (\rho_g \mathbf{u} Y_i)
      =  \nabla \cdot (\rho_g D \nabla Y_i)
      + S_Y
\end{equation}

$Y_i$ is the mass fraction of $i$-specie and $S_Y$ is a sink term introduced to represent the mass transfer from the gas to the solid phase due to adsorption; will be detailed in section \ref{SY_ST}. The axial dispersion is modeled using an effective dispersion coefficient $D_L$, as the following: 

\begin{equation}
    D = 0.7 D_m + 0.5 \bar{u} d_p
\end{equation}

\begin{equation}\label{DmCO2}
D_{m_{{\text{CO}_2},{\text{He}}}} = \frac{10^{-3} \cdot T^{1.75} \cdot \sqrt{\frac{1}{M_{\text{CO}_2}} + \frac{1}{M_{\text{He}}}}}
{p \left( \Sigma_{v,\text{CO}_2}^{1/3} + \Sigma_{v,\text{He}}^{1/3} \right)^2}
\end{equation}

$T$ is the Absolute temperature of the gas mixture, $M_{{CO}_2}$, $M_{He}$ are the Molar masses of CO$_2$ and He, respectively, $\Sigma_{v,\text{CO}_2}$, $\Sigma_{v,\text{He}}$ represent the collisions between CO$_2$ and He molecules, i.e. $\Sigma_{v,CO2}=26.9$, $\Sigma_{v,He}=2.88$ \cite{welty2014fundamentals} and $p$ is the operating pressure inside the packed bed.

$D_m$ is the molecular diffusion, e.g. here defined for a carbon dioxide helium (CO$_2$-He) mixture, derived from the Chapman–Enskog theory \cite{welty2014fundamentals}. Interested in a macroscopic scale behavior in beds that are not of very high or very low diameter to length aspect ratios, and assuming homogeneous nano-pores inside spherical adsorbing beads, then diffusion inside beds of packed spherical beads can be thus assumed isotropic. Of course, it is note worthy that one can redefine the diffusion term as a second rank-tensor that can be more adapted to anisotropic beds and adsorbing materials (e.g. defining different diffusion coefficients in the radial and axial directions). The same analysis of course applies to the bed porosity $\varepsilon_p$, that can also be defined as a 2nd-rank tensor, e.g. to account for the slight variations in bed porosity close to the walls or the packed-bed envelop.

\subsubsection{Energy conservation}
The energy conservation equation governs the thermal behavior of the coupled gas-solid system during the adsorption process, and is given by: 

\begin{equation} \label{equ:energyConservation}
    \frac{\partial}{\partial t} \left( \rho_{\text{eff}} c_{p_\text{eff}} T \right)
    + \nabla \cdot \left( {\varepsilon_t} \rho c_{p_\text{eff}} \, \mathbf{u} T \right) 
    - \nabla \cdot \left( \kappa_{\text{eff}} \nabla T \right) 
    =
    S_T
\end{equation}

In a porous media such as a packed bed of adsorbent porous spherical particles, energy transport occurs through both the fluid (gas) and solid phases {where \( \varepsilon_t \) is the total porosity combining all the gas volumes outside and inside the solid material (particles)}. Equation {\ref{equ:energyConservation}} accounts for transient heat accumulation, convective and conductive heat transport, and thermal effects induced by adsorption inside the packed bed. The right-hand side of equation \ref{equ:energyConservation} includes a source term $S_T$ that introduce the heat released due to the isosteric heat of adsorption, i.e. $\text{CO}_2$ adsorption on Zeolite-13X beads. $S_T$ will be detailed in section \ref{SY_ST}. The subscript "eff" denotes the effective thermophysical properties that are defined as volume-averaged quantities over the total porosity $\varepsilon_t$, thus blending {fluid and solid contributions \cite{zou2025fractal},} and of course assuming thermal equilibrium at the solid-gas interface . This simplification enables the treatment of the porous medium as homogeneous. 

Additionally, $\rho_{\text{eff}}$ is the effective density that reflects the weighted average of gas and solid mass per unit volume, defined as: 

\begin{equation}
      \rho_{\text{eff}} = {\varepsilon_b} \rho_g + \left( 1- {\varepsilon_b} \right) \rho_p
\end{equation}

$\rho_g$ is the gas density, $\varepsilon_b$ the bed bulk porosity and $\rho_p$ is the adsorbent particle density.

$c_{p_\text{eff}}$ is the effective heat capacity accounts for the heat storage capacity of both phases, defined as:

\begin{equation}
{
      c_{p_\text{eff}} = {\varepsilon_b} \frac{\rho_g}{\rho_{eff}} c_p + \left( 1- {\varepsilon_b} \right) \frac{\rho_p}{\rho_{eff}} c_{p_\text{p}}
      }
\end{equation}

$c_p$ is the gas heat capacity and $c_{p_\text{p}}$ the solid particle heat capacity.

$\kappa_{\text{eff}}$ is the effective thermal conductivity in $W\cdot m^{-1}K^{-1}$, defined as:

\begin{equation}
      \kappa_{\text{eff}} = {\varepsilon_b} \kappa + \left( 1- {\varepsilon_b} \right) \kappa_p
\end{equation}

Where $\kappa$ is the gas thermal conductivity and $\kappa_p$ the solid skeletal particles thermal conductivity that depends on the gas loading which is thus dependent of gas feed-in concentration or mass fraction; i.e. $\kappa_p=-0.331 (Y_{CO2-feed}) + 0.355$.

\subsection{An enhanced modeling of volumetric source terms}\label{SY_ST}

Many 1D, 2D and 3D CFD models from the literature \cite{HWANG1995,ramos2024cfd}, adopts the mathematical formulation of $S_Y$ (equation \ref{equ:speciesConservation}) and $S_T$ (equation \ref{equ:energyConservation}) as the following:

\begin{equation}\label{eqns_old_form_SY_ST}
\begin{array}{cc}
S_Y =  -\frac{(1- \varepsilon_b)}{\varepsilon_b} \rho_p \,  \sum_{i=1}^{n_{\text{comp}}} 
\left(M_i \frac{\partial q_i}{\partial t}\right)\\
S_T =  \frac{(1- \varepsilon_b)}{\varepsilon_b} \rho_p \,  \sum_{i=1}^{n_{\text{comp}}} 
\left(|\Delta H_i| \frac{\partial q_i}{\partial t}\right)    
\end{array}
\end{equation}

The mathematical form in both $S_Y$ and $S_T$ (equations \ref{eqns_old_form_SY_ST}), assume that authors adopted thus a unique volumetric source, dependent on $\frac{\partial q_i}{\partial t}$, that is responsible for both: (i) the gaseous i$^{th}$-component mass-fraction removal and (ii) the exothermic heat release; from the effective porous phase material of bulk porosity ($1-\varepsilon_b$) to the bulk gas mixture phase of bulk porosity $\varepsilon_b$. Though, mathematical modeling form of $S_Y$ and $S_T$ is true, it merits indeed some discussions and can be also subject to further enhancement. In fact, the form of $S_Y$ and $S_T$ in equation (\ref{eqns_old_form_SY_ST}) can not explicitly account for the adsorbing particle's porosity  $\varepsilon_p$ which is not necessarily equal to the bed's bulk porosity $\varepsilon_b$. Additionally, the mathematical form of $S_Y$ and $S_T$ (equation \ref{eqns_old_form_SY_ST}) can not account for the PAOR or gas loading inside the beads/pellets or particles. In other word they can not account for non-uniform $q_i$ inside the particles that will depend of course on the gas loading. The PAOR is a driving force behind the exothermic heat release and gas mass-fraction exchange at the gas-solid interface.

In order to enhance the physics-based background of the CFD modeling, one thus can account explicitly for the particle's porosity $\varepsilon_p$ and for the total porosity $\varepsilon_t$ in the source terms $S_Y$ and $S_T$, and can also account for the PAOR inside the beads/pellets. 

Agreeing on the above, for the first time, we thus propose a new more rigorous physics-based mathematical formulation of the two volumetric source terms $S_Y$ and $S_T$ as the following:

\begin{equation} \label{equ:DeltaYspeciesConservation}
      S_Y =  - \left[ (1-\varepsilon_p){\Gamma_Y}\right] \, \frac{(1- \varepsilon_t)}{\varepsilon_t} \rho_p \sum_{i=1}^{n_{\text{comp}}} 
    \left(M_i \frac{\partial q_i}{\partial t}\right)
\end{equation}

\begin{equation} \label{equ:DeltaTenergyConservation}
    S_T = \left[ (1-\varepsilon_p){\Gamma_T}\right] \, \frac{(1 - \varepsilon_b)}{\varepsilon_b} \rho_p \sum_{i=1}^{n_{\text{comp}}} \left( - \Delta H_i \frac{\partial q_i}{\partial t} \right)
\end{equation}

\begin{figure}
    \centering
    \includegraphics[width=0.45\textwidth]{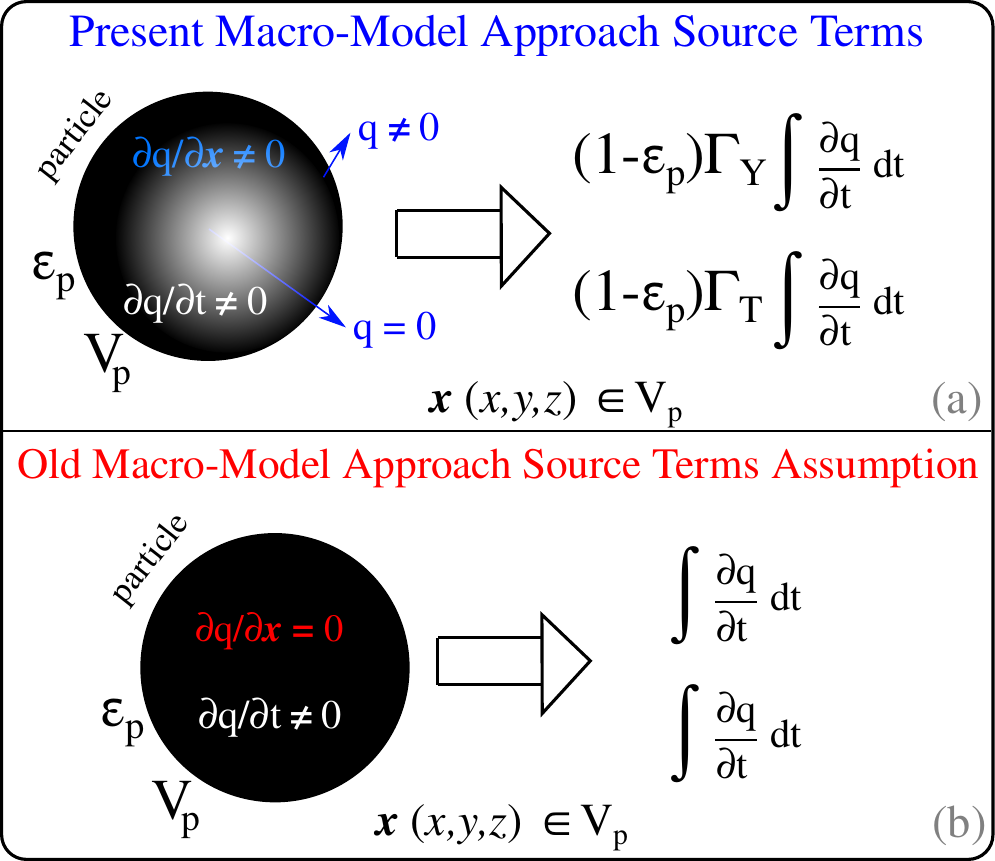}
    \caption{A schematic representation of the effect of the new introduced terms $(1-\varepsilon_p)\Gamma_Y$ of equation \ref{equ:DeltaYspeciesConservation} and $(1-\varepsilon_p)\Gamma_T$ of equation \ref{equ:DeltaTenergyConservation}. (a) Present two source terms modeling take into account the impact of PAOR (Pores Adsorption Occupation Rate) inside the adsorbing particles or beads/pellets in the new present 3D CFD model thus assuming ${{\partial q} \over {\partial \bf{x}}} \neq 0 $. (b) The two terms assumption in old macro-model formulations as presented in the literature that assume ${{\partial q} \over {\partial \bf{x}}}=0$. $q$ is the adsorbed quantity inside the particle of volume $V_p$ and porosity $\varepsilon_p$.}
    \label{fig:xi_Y_xi_T}
\end{figure}

 Here, \( \varepsilon_t \) is the total porosity, \( \varepsilon_p \) the particle's porosity, \( \varepsilon_b \) is the bed's bulk porosity (see figure \ref{fig:xi_Y_xi_T}),  \( \rho_p \) the adsorbent particle density, \( M_i \) the molar mass of gas specie \( i \), and \( q_i \) is the amount of specie \( i \) adsorbed per unit mass of adsorbent.

 One can see that in the new form of $S_Y$ (equation \ref{equ:DeltaYspeciesConservation}) and $S_T$ (equation \ref{equ:DeltaTenergyConservation}), whenever $\varepsilon_p=1$ or $\varepsilon_t=1$ then both $S_Y=0$ and $S_T=0$ which still holds true, because at either $\varepsilon_p=1$ or $\varepsilon_t=1$ then there should not exist any adsorbing bead/pellet material inside the bed. Moreover, with the new two terms $(1-\varepsilon_p){\Gamma_Y}$ and $(1-\varepsilon_p){\Gamma_T}$ one can now account for the impact of the PAOR inside the beads/pellets, which indeed depends on the initial CO\textsubscript{2} feed-in concentration (see figure \ref{fig:xi_Y_xi_T}).

\begin{figure*}
    \centering
        \includegraphics[width=0.9\textwidth]{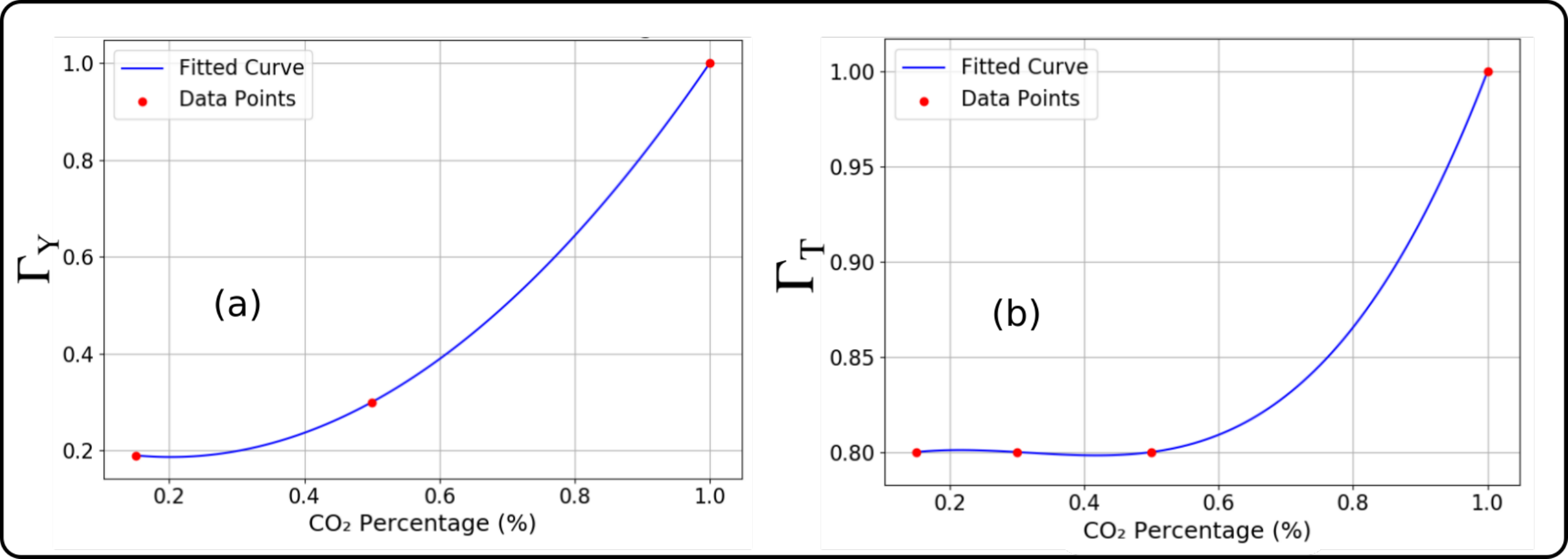}
    \caption{New volumetric source terms ($\Gamma_Y$ and $\Gamma_T$) as function of the CO\textsubscript{2} feed-in percentage. (a) $\Gamma_Y$; (b) $\Gamma_T$.}
    \label{fig:delta_fits}
\end{figure*}

Due to the direct impact of the gas feed-in percentage on the gas loading equilibrium, the values of $\Gamma_Y$ and $\Gamma_T$ can be thus correlated as function of the gas mixture feed-in concentrations, i.e. here is CO\textsubscript{2}. In the present work, this is done through comparisons with the experimental breakthrough curves and bed temperature profiles from the experiment by Wilkins et al. 2019 \cite{wilkins2019measurement}. Figure \ref{fig:delta_fits} shows an example of $\Gamma_Y$ and $\Gamma_T$ as function of the initial feed-in CO\textsubscript{2} percentage. The increase in $\Gamma_Y$ values (figure \ref{fig:delta_fits}-a) after a critical CO\textsubscript{2} feed-in threshold value of about 0.25 can be explained as the following: for values of feed-in of CO\textsubscript{2} $\leq 0.25$, $\Gamma_Y$ is constant and equals to about 0.2 meaning that 20\% of the total adsorbed quantity in the particles account for the adsorbed species mass-fraction exchange from the solid to the gas phase. While when the feed-in of CO\textsubscript{2} $> 0.25$, $\Gamma_Y$ values increase allowing thus for more species mass-fraction exchange between the solid and the gas phases. The increase in $\Gamma_T$ values (figure \ref{fig:delta_fits}-b) after a critical CO\textsubscript{2} feed-in threshold value of about 0.5 can be explained as the following: for values of feed-in of CO\textsubscript{2} $\leq 0.5$, $\Gamma_T$ is constant and equals to 0.8 meaning that 80\% of the total adsorbed quantity in the particles account for the heat release from the solid to the gas phase. While when the feed-in of CO\textsubscript{2} $> 0.5$, $\Gamma_T$ values increase allowing thus for more exothermic release of heat from the solid to the gas phase. 

{ The need for this modification of source terms is explained in details in figure \ref{fig:xi_Y_xi_T} where our new model of the source terms can take into account the impact of PAOR (pores adsorption occupation rate). This was never previously considered in any other CFD adsorption model in the literature, where authors usually assume a unique adsorpion rate quantity $\sum_{i=1}^{n_{\text{comp}}} \left(\frac{\partial q_i}{\partial t}\right)$ by the particles, and not a partial adsorpion rate quantity, $\Gamma \cdot \sum_{i=1}^{n_{\text{comp}}} \left(\frac{\partial q_i}{\partial t}\right)$ by the particles that can depend on the PAOR, thus the initial gas loading, where $\Gamma \leq 1 $.} 

It is also worth noting that many previous CFD models, implicitly assumed like $\Gamma_T= \Gamma_Y = 1$ for all CO\textsubscript{2} feed-in concentrations, which is not always necessarily true. This is because the feed-in concentration value usually has an impact in some way or another on the PAOR inside the beads/pellets.

\subsection{Adsorption kinetics and equilibrium}\label{AdsEquil}

The modeling of CO\textsubscript{2} adsorption relies on both equilibrium and kinetic considerations. For the equilibrium aspect, the dual-site Langmuir (DSL) isotherm model is employed \cite{glueckauf1955theory}, due to its capability to represent adsorption on heterogeneous surfaces through two distinct adsorption mechanisms. This model has been shown to effectively describe single-component adsorption on materials with energetically diverse sites \cite{keller2005gas}. The DSL model is represented by the following equation:

\begin{equation}
    q_i^* = \frac{ q_{s b, i} b_i c_i }{1 + \sum\limits_{i=1}^{n_{comp}} b_i c_i } + \frac{ q_{s d, i} d_i c_i }{1 + \sum\limits_{i=1}^{n_{comp}} d_i c_i}
\end{equation}

The different isotherm parameters are determined experimentally by Wilkins et al. 2019 \cite{wilkins2019measurement} and are used here for adsorption equilibrium modelling (see Table \ref{Tab:DSL}).

\begin{table}
\centering
\caption{Adsorption equilibrium isotherm data for CO\textsubscript{2} adsorption on Zeolite-13X, as adopted in the present 3D CFD modelling. Data as reported in the literature through experimental measurements by Wilkins et al. 2019 \cite{wilkins2019measurement}.}

\renewcommand{\arraystretch}{1.3}
\begin{tabular}{|c|c|c|c|}
\hline
\textbf{Symbol} & \textbf{Unit} & \textbf{Definition} & \textbf{Value} \\
\hline
\( q_{b,i}^{\text{sat}} \) & mol/kg & Saturation capacity of site \(b\) & 3.257 \\
\( b_0 \) & m\textsuperscript{3}/mol & Pre-exponential constant for site \(b\) & \(2.09 \cdot 10^{-7}\) \\
\( \Delta H_b \) & J/mol & Heat of adsorption on site \(b\) & -42670 \\
\( q_{d,i}^{\text{sat}} \) & mol/kg & Saturation capacity of site \(d\) & 3.240 \\
\( d_0 \) & m\textsuperscript{3}/mol & Pre-exponential constant for site \(d\) & \(1.06 \cdot 10^{-7}\) \\
\( \Delta H_d \) & J/mol & Heat of adsorption on site \(d\) & -32210 \\
\hline
\end{tabular}
\label{Tab:DSL}
\end{table}

The coefficients $b_i$ and $d_i$, representing the Langmuir affinity constants of component $i$ for the first and second adsorption site types, respectively, are determined using the following Arrhenius-type expressions, as reported by Haghpanah et al. (2013) \cite{haghpanah2013multiobjective}:
\begin{equation}
    b_i = b_0 \exp \left({\frac{- \Delta H_b}{RT}} \right)
\end{equation}

\begin{equation}
    d_i = d_0 \exp \left( {\frac{- \Delta H_d}{RT}} \right)
\end{equation}

The molar concentration $c_i$ is determined from the mass fractions $Y_i$ using the following relationship:
\begin{equation}
    c_i = \frac{P}{RT} \frac{\frac{Y_i}{M_i}}{\sum_{i=1}^{n}\frac{Y_j}{M_j}}
\end{equation}

From a kinetic standpoint, the linear driving force (LDF) model is adopted to characterize the rate of mass transfer. The LDF model is widely used for its simplicity and computational efficiency, enabling robust and streamlined predictions of adsorption dynamics without compromising the accuracy of the overall simulation.
\begin{equation}
      \frac{dq_i}{dt} = k_i(q^*_i - q_i)
\end{equation}
where $k_i$ is the LDF mass transfer coefficient for the i$^{th}$ gas component, $q^*_i$ is the equilibrium adsorbed amount and $q_i$ is the actual adsorbed amount i$^{th}$ gas component. The LDF mass transfer coefficient $k_i$ is usually obtained by simple experimental data fitting to the model equations \cite{Naidu2021}. However, a generalized model of $k_i$ is usually employed which depends on the molecular diffusion of the i$^{th}$ gas component in the gaseous mixture, and that is inversely proportional to the diameter $d_p$ of the adsorbing particles such that:

\begin{equation}
{
    k_i = \frac{15\varepsilon_p}{r^2_p}  \frac{D_m}{\tau} \frac{c_i}{q^*_i}
}
\end{equation}

{Where $\tau$ is the turtuosity considered equal to 3 (see \cite{wilkins2019measurement}).} For different macropore models of $k_i$ depending on the type and form of adsorbing particles, the reader may refer to table 1 in Rezaei et al. 2009 \cite{REZAEI2009}. In the present work, we adopt the $k_i$ model intended for spherical beads/pellets (see \cite{REZAEI2009}).

\subsection{Initial and Boundary Conditions} \label{Initial and Bondary Conditions}

To investigate the influence of gas composition on the adsorption and heat transfer behavior in the fixed packed-bed column, 3D CFD transient simulations were carried out to predict the breakthrough curves in a cylindrical fixed-bed adsorber experiment by Wilkins et al. 2019 \cite{wilkins2019measurement}. Three feed-in gas mixture scenarios are computationally investigated: 100\% CO\textsubscript{2}, 50\% CO\textsubscript{2} / 50\% He, and 15\% CO\textsubscript{2} / 85\% He. Each configuration was applied with the appropriate corresponding volumetric flow rate, reflecting the real experimental operating conditions. The initial pressure is kept constant at 102000 Pa for all cases as in the experiment by Wilkins et al. 2019 \cite{wilkins2019measurement}. Table \ref{tab:inlet_conditions} summarizes the inlet feed-in gas mixture conditions used in the present 3D CFD simulations.

\begin{table}
\centering
\caption{Present 3D CFD feed-in conditions. Gas mixture composition and flow rates with data as reported in the literature through experimental measurements by Wilkins et al. 2019 \cite{wilkins2019measurement}.}
\begin{tabular}{|l|l|l|l|}
\hline
\textbf{Gas Mixture} & \textbf{CO\textsubscript{2} (\%)} & \textbf{He (\%)} & \textbf{Volumetric Flow Rate (${m^3}/s$)} \\ \hline
\midrule
(CO\textsubscript{2})      & 100 & 0   & $5.83 \cdot 10^{-6}$ \\
(CO\textsubscript{2} , He)  & 50  & 50  & $5.25 \cdot 10^{-6}$  \\
(CO\textsubscript{2} , He)  & 15  & 85  & $11 \cdot 10^{-6}$     \\
\hline
\end{tabular}\label{tab:inlet_conditions}
\end{table}

The 3D CFD simulation is initialized with uniform fields for the velocity, pressure, and temperature. At the bed's inlet a constant volumetric flow rate of the CO\textsubscript{2}-N\textsubscript{2} mixture is imposed, while the outlet is modeled using a total outlet pressure condition. For the bed's wall boundaries, a convective heat transfer model is applied to appropriately account for the natural convection heat transfer. Table~\ref{tab:boundaryConditions} summarizes the boundary conditions used in the present 3D CFD transient simulations.

\begin{table}
\centering
\caption{Initial and boundary condition (BC) applied in the present transient 3D CFD model.}
\begin{tabular}{|l|l|l|l|}
\hline
\textbf{Field} & \textbf{Region} & \textbf{BC Type} & \textbf{Value} \\ 
\hline
\multirow{3}{*}{Pressure $p$} & Internal & \textit{Dirichlet} & 1.02 {bar} \\
 & Inlet & \textit{Neumann} & -- \\
 & Outlet & \textit{Dirichlet} & 1.02 {bar} \\
 & Wall & \textit{Neumann} & -- \\
\hline
\multirow{3}{*}{Velocity $\mathbf{U}$} & Internal & \textit{Dirichlet} & $\mathbf{U}=0$ \\
 & Inlet & \textit{Dirichlet} (Flow Rate)  & table \ref{tab:inlet_conditions} \\
 & Outlet & \textit{Neumann} & -- \\
 & Wall & \textit{Dirichlet} (No slip) & $\mathbf{U}=0$ \\
\hline
\multirow{3}{*}{Temperature $T$} & Internal & \textit{Dirichlet} & 294.6 K \\
 & Inlet & \textit{Dirichlet} & 294.6 K \\
 & Outlet & \textit{Neumann} & -- \\
 & Wall & \textit{Neumann} & eqn. \ref{eqn_h} \\
\hline
\end{tabular}
\label{tab:boundaryConditions}
\end{table}

The external wall heat transfer is modeled through a natural convection correlation implemented as a coded boundary condition in \href{https://www.openfoam.com}{\textbf{OpenFOAM}}. This approach allows the convective heat transfer coefficient to adapt dynamically based on the height/position of the cylindrical bed and the maximum temperature at each time step. This heat transfer model is based on empirical correlations for natural convection, (see Perry’s Chemical Engineers’ Handbook (2008) \cite{doble1984perry}). For high Rayleigh number flows ($\text{Gr} \cdot \text{Pr} > 10^9$), the local heat transfer coefficient, can be expressed as:

\begin{equation}\label{eqn_h}
    h(z,t) = 1.24 \left[ T(z,t) - T_{ref} \right]^{1/3}
\end{equation}

where $Gr$ is the Grashof number, $Pr$ the Prandtl number, $h(z,t)$ the convective heat transfer coefficient in $W \cdot m^{-2} \cdot K^{-1}$, $T(z,t)$ the temperature of the wall at each time step and position z, and $T_{ref}$ is the reference ambient temperature, typically taken as ${294~K}$. This formulation is thus appropriate to account for natural convection heat transfer that is both local-position and temperature-dependent. This allows for a more physically accurate representation of thermal exchange between the bed's cylindrical wall and the surrounding environment (air). This boundary treatment enhances the accuracy of wall heat transfer modeling, especially in systems dominated by natural convection, such as in fixed-bed adsorption columns with no exterior thermal insulation treatment.

\section{Numerical Methodology}
\label{sec:num_method}

The 3D CFD transient simulations are conducted within the open-source CFD platform \href{https://www.openfoam.com}{\textbf{OpenFOAM}}, version 2306. An in-house transient solver is developed to resolve the coupled transport equations for mass, momentum, energy, and species in a porous medium undergoing gas adsorption. The finite volume method (FVM) is employed for spatial discretization, and time integration is carried out using a first-order implicit scheme to ensure stability and allow large time steps during transient simulations. 

The fixed column cylindrical packed bed is treated as a continuum porous medium, with its impact on fluid flow incorporated through porous-based volume-averaged momentum source term. These include both viscous and inertial resistances, implemented using a \texttt{Darcy-Forchheimer} model formulation available via an \texttt{explicitPorositySource} term. The solid structure is defined over a designated cell zone denoted \texttt{porosity}, where the porosity model is activated.

To characterize the resistance imposed by the porous matrix in the packed bed, the CFD model applies volume-averaged properties based on a defined constant initial bed porosity and particle's or pellet's diameter. A summary of the cylindrical fixed-bed properties and adsorbent Zeolite-13X material properties are provided in Table~\ref{tab:column_properties}.

\begin{table}
\centering
\caption{Fixed-bed and adsorbent material properties in the present 3D CFD simulations.}
\begin{tabular}{|l|l|}
\hline
\textbf{Fixed-bed Properties} & \textbf{Value} \\
\hline
Column length, $L$ & 0.064 $m$ \\
Column inner, radius $r$ & 0.0141 $m$ \\
Bed porosity, $\varepsilon_b$ & 0.48 \\
Particle porosity, $\varepsilon_p$ & 0.35 \\
\hline
\textbf{Material Properties (Zeolite 13X)} & \\
\hline
Particle diameter, $d_p$ & 1 $mm$ \\
Particle density, $\rho_p$ & 1050 $kg/ m^{3}$ \\
Specific heat capacity, $c_{p}$ & 856 $J/kg/K$ \\
Thermal conductivity, $\kappa_p$ &  { $\kappa_0$ + $\beta_0$ ($T-T_{ref}$) \cite{vyas1995estimation} } \\
{} & see table \ref{tab:kappa_params}\\
\hline
\end{tabular}
\label{tab:column_properties}
\end{table}

The thermophysical properties of the different involved gaseous species, such as density, specific heat capacity, thermal conductivity, and viscosity, are all calculated an updated iteratively in the solver using the thermodynamics gas library data modules available in \href{https://www.openfoam.com}{\textbf{OpenFOAM}} (e.g. NIST thermodynamics tables: Thermophysical Properties of Fluid Systems).  The viscous and inertial coefficients in the Darcy-Forchheimer model are determined based on the Ergun correlation, as expressed in equation (\ref{equ:ergun}). The flow resistance is thereby introduced explicitly in the momentum equation, which ensures a proper pressure drop and velocity attenuation due to the presence of the packed adsorbent material zeolite-13X beads. This approach preserves computational efficiency while capturing the essential characteristics of flow through an isotropic porous media.

\subsection{Geometry and Mesh Generation}

\begin{figure}
    \centering
\includegraphics[width=0.48\textwidth]{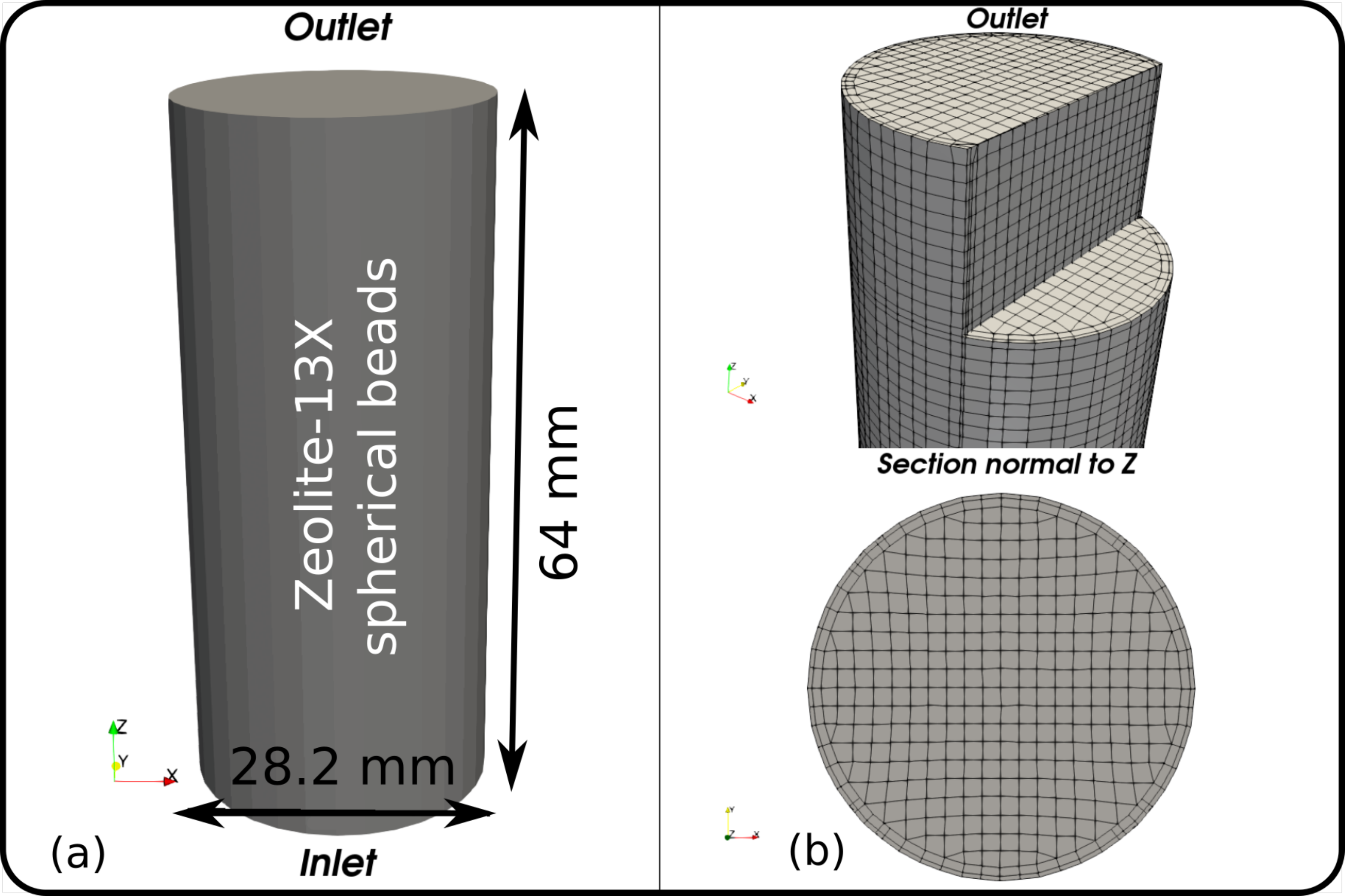}
    \caption{{A perspective view of the 3D CFD computational domain of the fixed-bed adsorber. (a) bed dimensions; (b) the mesh topology with boundary mesh layers close to the external wall of the bed.}}
    \label{fig:Mesh_Cyl}
\end{figure}

The computational domain is made of a cylindrical packed bed adsorber with a diameter of 28.2 $mm$ and a height of 64 $mm$ as shown in figure \ref{fig:Mesh_Cyl} (see \cite{wilkins2019measurement}. The mesh is generated in two stages: a structured background mesh is created using \texttt{blockMesh}, followed by geometry snapping and cells refinement using the \texttt{snappyHexMesh} mesher tool in \href{https://www.openfoam.com}{\textbf{OpenFOAM}}. Boundary mesh layers are then added close to the bed wall surfaces to improve the resolution of the thermal and velocity boundary layers, as shown in Figure \ref{fig:Mesh_Cyl}. 

\begin{table}
    \centering
    \caption{Mesh configurations and corresponding values of the Grid Convergence Index (GCI) \cite{roache1994perspective,Celik2008}. An example based on $T$ values at the time instance of $t = 120$~s.}
    \label{tab:gci_results}
    \begin{tabular}{|c|c|c|}
        \hline
        \textbf{Mesh Type} & \textbf{Number of Cells} & \textbf{GCI (\%)} \\
        \hline
        Fine (F)     & 800000   & Base       \\
        Medium (M)   & 337500   & 0.24 (M-F) \\
        Coarse (C)   & 100000   & 0.27 (C-M) \\
        \hline
    \end{tabular}
\end{table}

A grid convergence study is performed to ensure that the CFD numerical results are insensitive to the mesh-cell size being adopted. Table~\ref{tab:gci_results} shows the mesh configurations used in the grid convergence study, adopting the corresponding Grid Convergence Index (GCI) approach \cite{roache1994perspective,Celik2008}. Three levels of mesh resolution were tested, coarse (100,000 cells), medium (337500 cells), and fine (800000 cells), to assess the numerical sensitivity of the temperature field at t=120 s. The GCI, which quantifies the relative error between successive mesh refinements, was calculated between the coarse-to-medium and medium-to-fine mesh pairs \cite{roache1994perspective,Celik2008}. The resulting Medium-Coarse mesh-size GCI \% value of 0.27\% (table \ref{tab:gci_results}) indicates that the coarse mesh of 100000 cells can be confidently adopted. It provides a sufficient numerical accuracy with significantly reduced computational cost compared to the fine mesh. An initial time step of 4e-4 $s$ is used but with adaptive time-step adjustment in the solver based on the Courant–Friedrichs–Lewy number (CFL) condition (CFL $< 1$). This ensures both computational speed along with good accuracy and numerical stability. Convergence criteria at each time step are based on the residuals of pressure and velocity thresholds that are set to very low values of $10^{-6}$ and $10^{-7}$, respectively.

\section{Numerical Validation with Experiments}
\label{sec:validation}

To assess the accuracy and predictive capability of the newly developed 3D CFD model, the numerical results are compared to the experimental measurements data reported by Wilkins et al. 2019 \cite{wilkins2019measurement} and to the 2D CFD results of Ramos et al. (2024)~\cite{ramos2024cfd}. The validation process is structured into two distinct stages. The first focuses on pure CO\textsubscript{2} adsorption conditions, allowing evaluation of the model's ability to capture single-species transport, thermal dynamics, and adsorption behavior. The second stage extends the analysis to multi-component inlet conditions, specifically mixtures containing 15\% and 50\% CO\textsubscript{2} in He, in order to evaluate the model’s performance in simulating competitive adsorption and multi-species mass transfer.

In the validation phase, particular attention is given to several critical indicators that characterize the dynamic behavior of the adsorption process. One of the primary quantities of interest is the temperature peak-value observed locally inside the packed bed, which arises due to the exothermic nature of adsorption in Zeolite-13X beads. Accurately capturing the location and magnitude of this temperature maximum is essential for validating the thermal coupling with adsorption kinetics implemented in the present 3D CFD model.

To ensure consistency and relevance in the comparisons, the temperature values are extracted at a probe located at the same position defined in the experimental study by Wilkins et al. 2019 \cite{wilkins2019measurement} at the centerline of the cylindrical bed center-line ($r=0$) and at a height of $z = 52~mm$ away form the feed-in inlet. The temperature profiles along this probe serve as the primary metric for validation with the experimental data Wilkins et al. 2019 \cite{wilkins2019measurement}, and with the 2D CFD data of Ramos et al. (2024)~\cite{ramos2024cfd}.

Additionally, the breakthrough time, defined as the point at which the adsorbent becomes saturated and the adsorbed species begins to appear at the outlet, is used as a key metric to assess the model's capacity to predict adsorption front propagation. This time marks the end of the effective adsorption cycle and is determined from both simulation results and experimental data. To facilitate quantitative comparisons, the outlet concentration of CO\textsubscript{2} is integrated over the outlet surface and compared against the experimentally observed breakthrough curves. This approach enables direct evaluation of the model’s accuracy in capturing the adsorption dynamics under different operating conditions, including pure CO\textsubscript{2} and binary mixtures with He at 15\% and 50\% inlet molar fractions of CO\textsubscript{2}.

\subsection{3D CFD Model Validation}

\subsubsection{Feed-in 100\% CO\textsubscript{2}}

To assess the fidelity of the proposed three-dimensional CFD model, a validation study was conducted using a case scenario involving the injection of a pure CO\textsubscript{2} stream (100\% molar fraction) into a fixed porous cylindrical adsorption bed. This scenario represents an idealized yet practically relevant condition for characterizing the adsorption behavior of CO\textsubscript{2} onto solid adsorbents, enabling clear interpretation of transport phenomena and thermal interactions in the absence of competing gas species, and facilitating direct comparison with experimental results reported by Ramos et al.~\cite{ramos2024cfd}.

Overall, the use of a pure CO\textsubscript{2} inlet stream not only maximizes the adsorption depth into the porous matrix but also provides the clearest correlation between the model’s outputs and experimental observables—namely, adsorption capacity and thermal response. This approach serves as the most effective basis for model validation, ensuring accurate representation of both mass and energy transport phenomena in subsequent simulations involving gas mixtures or cyclic adsorption processes. Under 100\% CO\textsubscript{2} inlet conditions, the adsorption front advances more uniformly and deeply into the bed compared to diluted mixtures, due to the higher partial pressure gradient driving mass transfer and the increased availability of adsorbate molecules. This facilitates the assessment of breakthrough dynamics, bed saturation zones, and localized heat release, offering robust data for model verification.

Furthermore, the adsorption capacity attained under pure CO\textsubscript{2} flow aligns closely with equilibrium data derived from experimentally measured adsorption isotherms conducted under similar conditions. These isotherms, typically obtained via volumetric or gravimetric methods, represent the maximum uptake achievable at specific temperatures and pressures. When applied to the CFD model, the isotherm data serve as boundary conditions or validation checkpoints, reinforcing the model's ability to predict adsorption behavior under saturation-limited scenarios.

\begin{table}
\centering
\caption{Thermal conductivity law physical parameters\cite{vyas1995estimation}.}
\label{tab:kappa_params}
\begin{tabular}{|c|c|c|}
\hline
\textbf{Parameter} & \textbf{Value} & \textbf{Unit} \\ 
\hline
$\kappa_0$ & $8.17635 \cdot 10^{-2}$ & $W \cdot m^{-1} \cdot K^{-1}$ \\
$\beta_0$ & $10.915427 \cdot 10^{-4}$ & $W \cdot m^{-1}\cdot K^{-2}$ \\
$T_{ref}$ & $303$ & K \\ 
\hline
\end{tabular}
\end{table}

\begin{figure}
\centering
\includegraphics[width=0.5\textwidth]{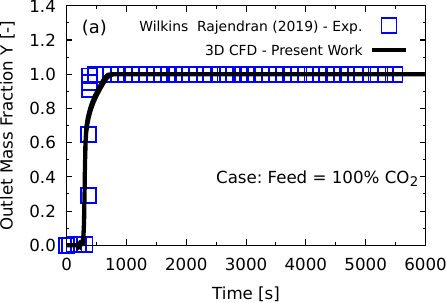}
\includegraphics[width=0.5\textwidth]{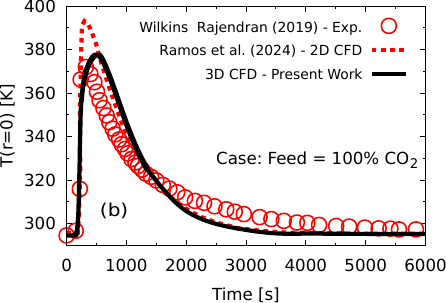}
\caption{{Validation of the present 3D CFD model results against the experimental data of Wilkins et al. 2019 \cite{wilkins2019measurement}. Case at 100\% CO\textsubscript{2} feed-in and $5.83 \cdot 10^{-6} m^3/s$ flow rate, $p=1.02$ bar. Temperature profile corresponds to the centerline local position ($r=0$) located at $z=52 mm$ away from the bed inlet. (a) Outlet CO\textsubscript{2} mass fraction (Y) as a function of time in seconds; and (b) local temperature T at $r=0$ and $z=52 mm$ inside the bed as function of time in seconds.}}
\label{fig:mass_temp_validation}
\end{figure}

Figure \ref{fig:mass_temp_validation}a presents the temporal evolution of the outlet CO\textsubscript{2} mass fraction. The present 3D CFD model accurately reproduces the initial delay in outlet concentration associated with the breakthrough curve, demonstrating its ability to correctly capture the mass transfer resistance and bed saturation behavior. The predicted breakthrough time closely matches the experimental value, indicating proper implementation of adsorption kinetics and transport limitations.

Simultaneously, thermal validation is achieved by comparing the axial centerline temperature rise predicted by the simulation to the experimentally measured temperature profile, as shown in Figure \ref{fig:mass_temp_validation}b. The model captures both the timing and magnitude of the temperature peak resulting from the exothermic nature of CO\textsubscript{2} adsorption. This agreement confirms the robustness of the energy conservation formulation, including the coupling between adsorption heat generation and thermal diffusion within the packed bed.

Overall, the pure CO\textsubscript{2} validation case demonstrates that the 3D CFD model reliably predicts both mass transport and heat release phenomena. It forms a foundational validation step prior to extending the analysis to more complex multicomponent feed conditions and cyclic adsorption scenarios.

\subsubsection{Feed-in 50\% CO\textsubscript{2} 50\% He Mixture}

\begin{figure}
    \centering
        \includegraphics[width=0.49\textwidth]{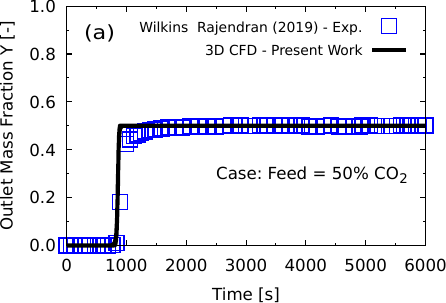}
        \includegraphics[width=0.49\textwidth]{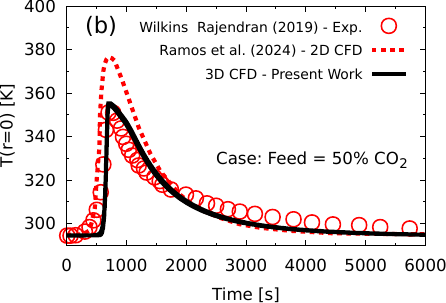}
    \caption{{Validation of the present 3D CFD model results against the experimental data of Wilkins et al. 2019  \cite{wilkins2019measurement}. Case at 50\% CO\textsubscript{2} 50\% He feed-in and $5.25 \cdot 10^{-6}~m^3/s$ flow rate, $p=1.02$ bar. Temperature profile corresponds to the centerline local position ($r=0$) located at $z=52~mm$ away from the bed inlet. (a) Outlet CO\textsubscript{2} mass fraction (Y) as a function of time in seconds; and (b) local temperature T at $r=0$ and $z=52~mm$ inside the bed as function of time in seconds.}
}
    \label{fig:mass_temp_validation_50}
\end{figure}

Figure~\ref{fig:mass_temp_validation_50} presents the validation of the 3D CFD model against experimental data for a 50\% CO\textsubscript{2} feed-in in CO\textsubscript{2}-He mixture at 1.02 bar. The outlet mass fraction of CO\textsubscript{2} shown in Figure \ref{fig:mass_temp_validation_50}a, demonstrates strong alignment with the experimental breakthrough curve. The onset and evolution of the breakthrough are accurately reproduced by the present 3D CFD model. 

The corresponding thermal response, illustrated in Figure \ref{fig:mass_temp_validation_50}b, confirms the model's ability to capture the temperature rise resulting from exothermic adsorption. The predicted peak temperature closely matches the experimental profile, with a deviation of $0.8$~K only. 

\begin{figure}
    \centering
        \includegraphics[width=0.499\textwidth]{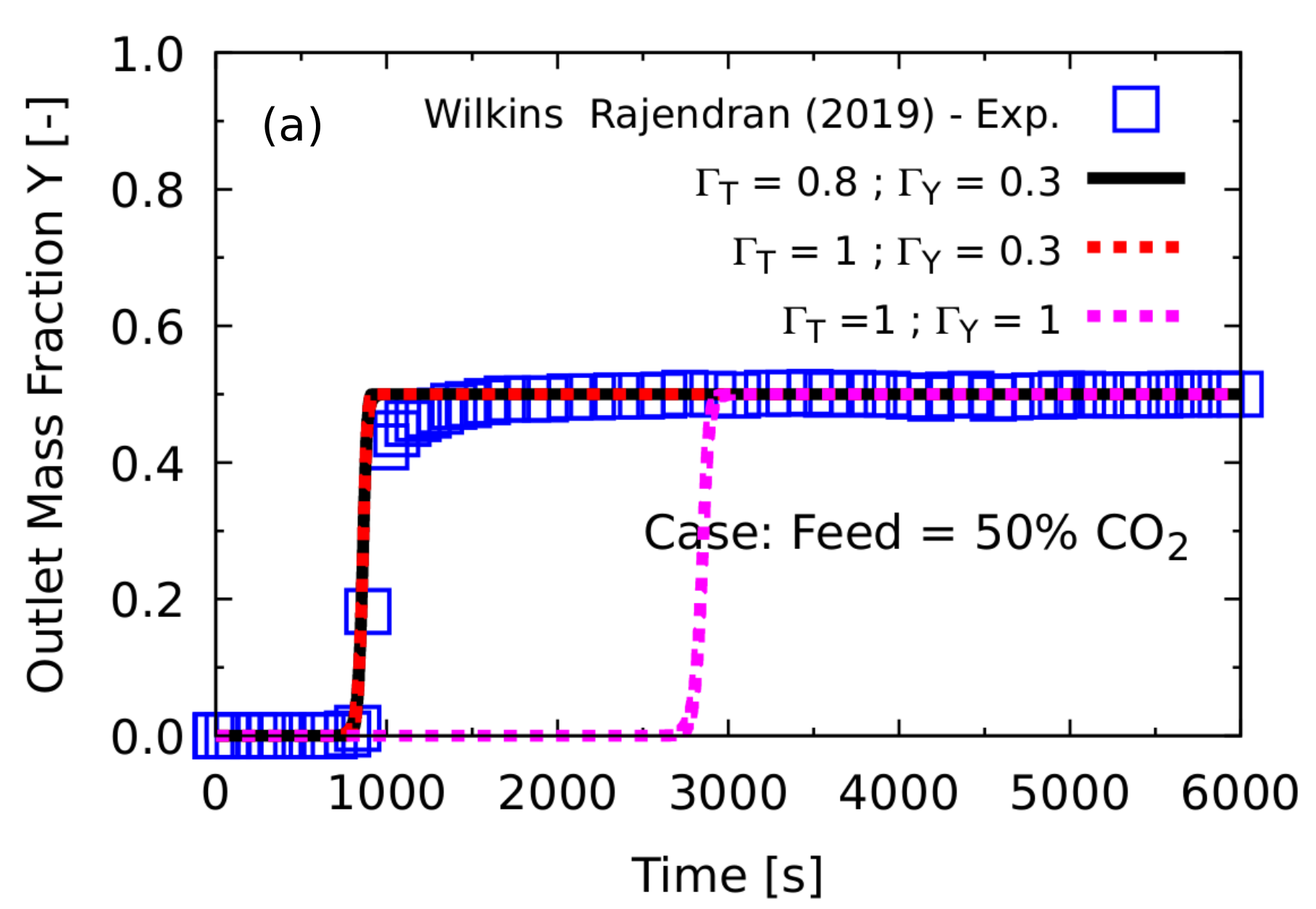}
        \includegraphics[width=0.49\textwidth]{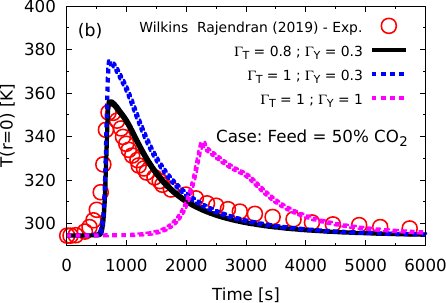}
    \caption{An example of the impact of the new terms $\Gamma_T$ and $\Gamma_Y$ representing the PAOR (Pores Adsorption Occupation Rate) inside the adsorbing particles or beads/pellets in the present new 3D CFD model. Case at 50\% CO\textsubscript{2} 50\% He feed-in and $5.25 \cdot 10^{-6}~m^3/s$ flow rate, $p=1.02$ bar. Temperature profile corresponds to the centerline local position ($r=0$) located at $z=52~mm$ away from the bed inlet. (a) Outlet CO\textsubscript{2} mass fraction (Y) as a function of time in seconds; and (b) local temperature T at $r=0$ and $z=52~mm$ inside the bed as function of time in seconds.}
    \label{fig:impact_Gamma}
\end{figure}

{Figure \ref{fig:impact_Gamma} shows an example of the impact of the implemented new terms $\Gamma_T$ and $\Gamma_Y$ on the results (see equation \ref{equ:DeltaYspeciesConservation} and equation \ref{equ:DeltaTenergyConservation}). Worth noting that these two terms represent the PAOR (Pores Adsorption Occupation Rate) inside the adsorbing particles or beads/pellets in the present new 3D CFD macro-scale model.}

\subsubsection{Feed-in 15\% CO\textsubscript{2} 85\% He Mixture}

\begin{figure}
    \centering
        \includegraphics[width=0.49\textwidth]{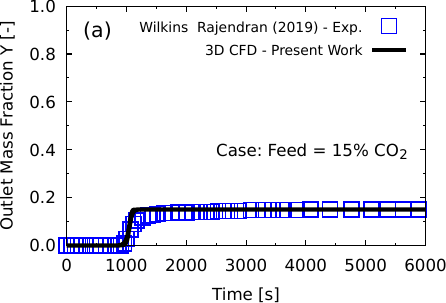}
        \includegraphics[width=0.49\textwidth]{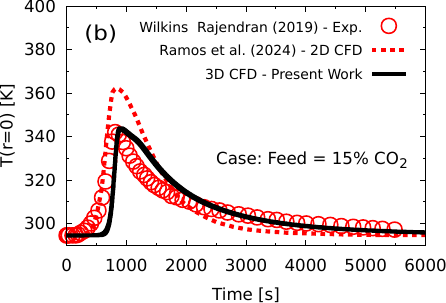}
    \caption{{Validation of the present 3D CFD model results against the experimental data of Wilkins et al. 2019  \cite{wilkins2019measurement}. Case at 15\% CO\textsubscript{2} 85\% He feed-in and $11 \cdot 10^{-6}~m^3/s$ flow rate, $p=1.02$ bar. Temperature profile corresponds to the centerline local position ($r=0$) located at $z=52~mm$ away from the bed inlet. (a) Outlet CO\textsubscript{2} mass fraction (Y) as a function of time in seconds; and (b) local temperature T at $r=0$ and $z=52~mm$ inside the bed as function of time in seconds.}}
    \label{fig:mass_temp_validation_15}
\end{figure}

Building on the previous case, figure \ref{fig:mass_temp_validation_15} shows the model validation for a more dilute feed containing 15\% CO\textsubscript{2} and 85\% He at 1.02 bar. As shown in figure \ref{fig:mass_temp_validation_15}a, the outlet mass fraction curve from the present 3D CFD simulation accurately predicts the observed experimental breakthrough curve.

Figure \ref{fig:mass_temp_validation_15}b displays the thermal response along the axial centerline (r=0) for 15\% CO$_2$ feed-in  CO$_2$-He mixture. The present 3D CFD model accurately predicts the experimental temperature evolution with high fidelity, including the timing and amplitude of the thermal peak. 

Overall, the present 3D CFD results confirm the reliability of the present 3D CFD model under different feed-in conditions. The consistent application of calibrated isotherms, combined with gas-specific thermal corrections, enables accurate prediction of both mass and energy transport phenomena inside the fix-bed column.

\subsection{3D CFD Results and Discussions}
\label{sec:results_discussion}

\begin{figure}
    \centering
\includegraphics[width=0.48\textwidth]{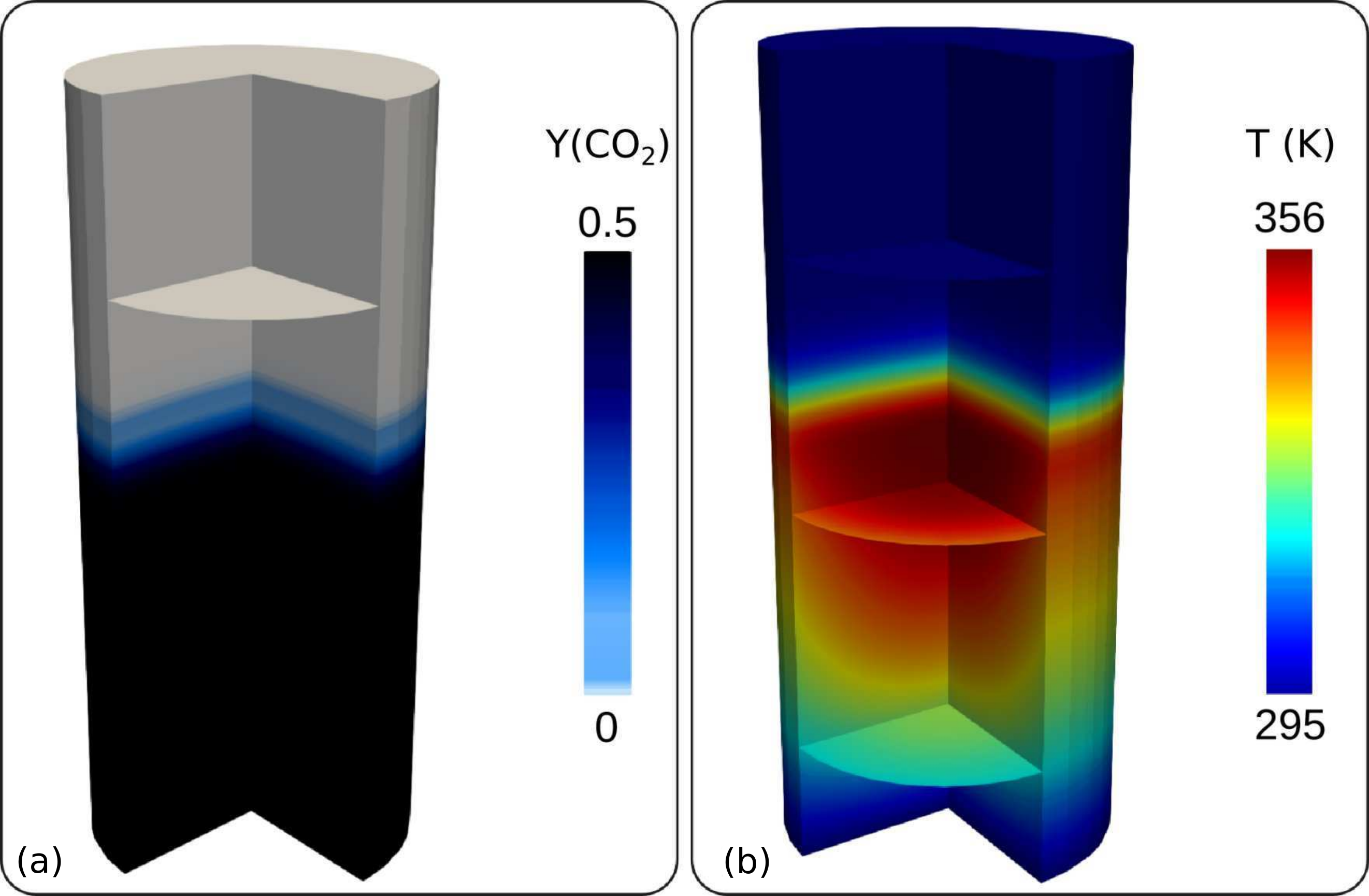}
    \caption{{Present 3D CFD results of adsorption inside a cylindrical packed bed of porous Zeolite-13X spherical beads. Results for 50\% CO2 50\% He feed-in gas mixture, showing: (a) the mass-fraction of CO2 concentration and (b) the temperature profile at time instance t = 500s.}}
    \label{fig:MeshCyl}
\end{figure}

To provide a spatially resolved visualization of the adsorption front dynamics inside the bed, Figure \ref{fig:MeshCyl} shows the 3D contours of CO\textsubscript{2} mass fraction and local temperature distribution within the porous bed at $t = 500$~s for the 50\% CO\textsubscript{2} 50\% He feed-in gas mixture case. These results highlight the interplay between mass transport and thermal effects during the adsorption process. The CO\textsubscript{2} mass fraction distribution, Figure \ref{fig:MeshCyl}-(a), clearly reveals the evolution of the adsorption front, with a sharp gradient delineating the boundary between the saturated and unsaturated zones in the adsorbing material. On the Figure \ref{fig:MeshCyl}-(b), the temperature field shows a pronounced thermal front resulting from the exothermic nature of the adsorption reaction, with maximum peak value in the central region of the bed where adsorption is taking place. These 3D insights complement the 1D validation plots and reinforce the model’s ability to capture the coupled mass and energy transport phenomena under realistic operating conditions.

\begin{figure}
    \centering
        \includegraphics[width=0.5\textwidth]{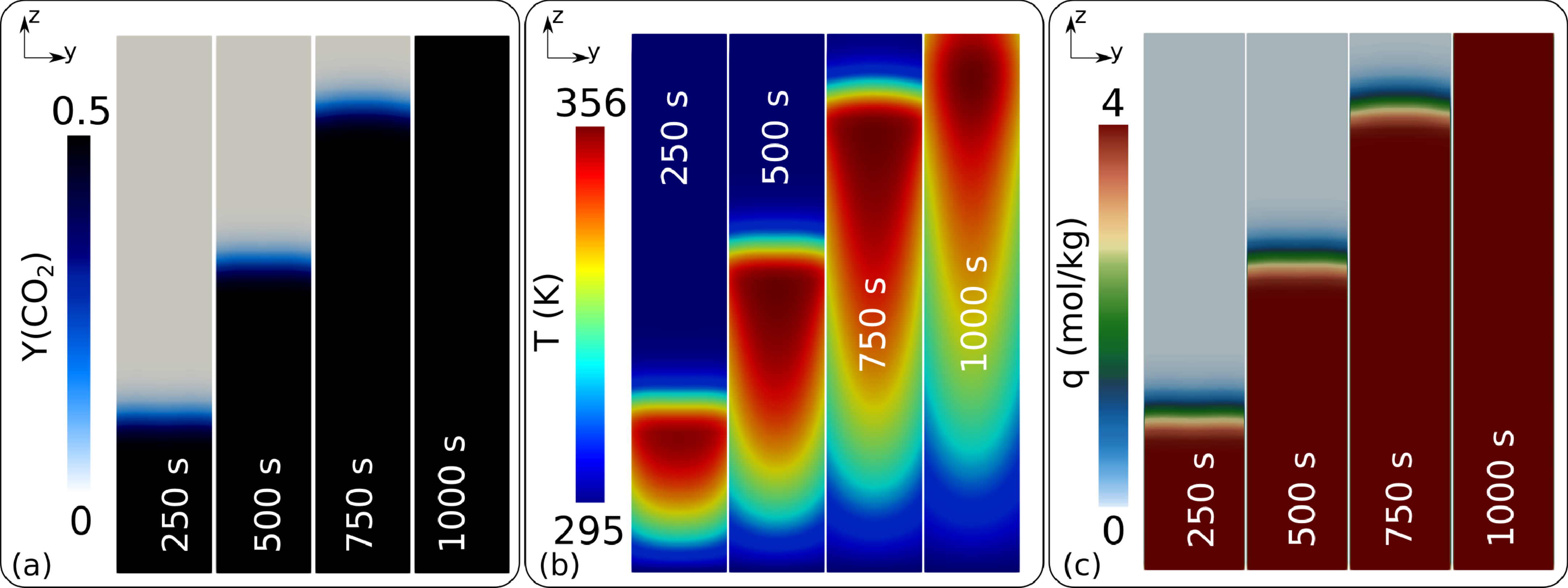}
    \caption{Time-resolved present CFD results for the cylindrical packed bed of porous Zeolite-13X spherical beads. (a) CO\textsubscript{2} mass fraction $(Y)$, (b) temperature $(T)$, and (c) adsorbed CO\textsubscript{2} quantity $q_0$ $(mol/kg)$ at respective time instances of 250 s, 500 s, 750 s and 1000 s.}
    \label{fig:comparison_time_fields_Cyl}
\end{figure}

Figure \ref{fig:comparison_time_fields_Cyl} shows an example of cross-sectional slices perpendicular to the x-axis of the cylinder bed that illustrate the local spatial distributions of CO\textsubscript{2} mass fraction, temperature, and the adsorbed CO\textsubscript{2} quantity $q_0$ inside the bed. As time progresses, the CO\textsubscript{2} breakthrough front advances along the column from inlet to outlet, accompanied by a sharp rise in the local temperature due to the exothermic nature of adsorption on Zeolite-13X spherical beads. The thermal maximum peak value displaces locally following the adsorption front with some phase-shift. The $q_0$ profile highlights the progressive saturation of the adsorbent material. For example, at t=1000 s, the adsorption front is near the bed's outlet, and the thermal front begins to dissipate, illustrating the onset of saturation and heat redistribution within the bed. These results emphasize the strong spatial and temporal coupling between mass and heat transport phenomena that underline the necessity of using a fully resolved 3D model to accurately capture such dynamics especially in future complex designs of adsorbing fixed-beds.

\section{CO2 Capture: A New Design of Fixed-bed Adsorber}

\begin{figure}
    \centering
    \includegraphics[width=0.48\textwidth]{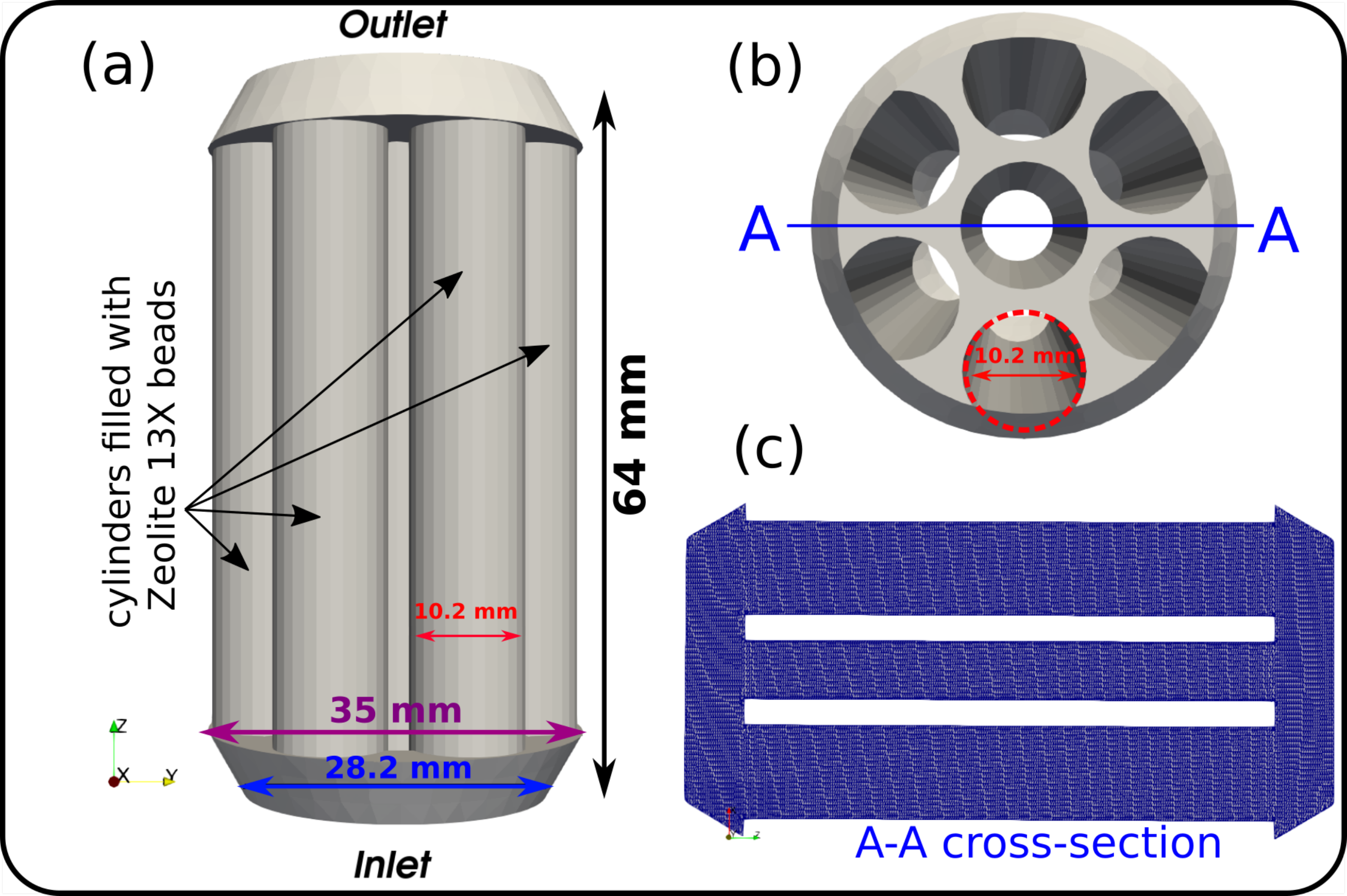}
    \caption{{A new design of fixed-bed adsorber made of seven inter-spaced cylindrical tubes that are connected between the inlet and outlet, $28.2~mm$ each, as same inlet/outlet of the reference bed design shown in figure \ref{fig:Mesh_Cyl}a. All the tubes are filled with the same Zeolite 13X beads as in the reference tube design shown in figure \ref{fig:Mesh_Cyl}a. (a) zy-plane perspective view showing bed dimensions in millimeters; (b) zy-plane top view; (c) cross section view.}
    }
    \label{fig:complex_geometry}
\end{figure}

While the preceding CFD results are obtained employing a single-cylinder adsorption bed, the full 3D capabilities of our solver enable computations within more intricate geometries or bed designs of complex geometries. To demonstrate this versatility, we extended our analysis to investigate a new design of adsorbing bed configuration presented in figure \ref{fig:complex_geometry}. This new bed design in figure \ref{fig:complex_geometry} have the same material and equal volume, thus the same quantity of Zeolite-13X solid adsorbent material as used in the initial cylindrical fix-bed design of figure \ref{fig:MeshCyl}a.

 {Multitubular adsorber configurations have been investigated by Carmo et al. (2020) \cite{carmo2020recovery} who developed a 2-dimensional mathematical model to sudy vinyl chloride recovery using TPSA system. The innovtion of such new design in figure 9 lies within its ability to reduce the maximum local peak temperature value, and to reduce the overall cooling period, without any additional cost (using same mass of adorbent material). This innovation is very beneficial in cyclic adsorption, because one can do with this design less number of total cycles compared to a one-single-cylindrical bed design. } 

The new design in figure \ref{fig:complex_geometry} significantly increases the external surface area, which enhances heat exchange with the surroundings. Note that such a configuration cannot be resolved using conventional 1D models or axisymmetric 2D CFD solver. The ability of our 3D solver to fully resolve these interactions allows for the exploration of novel bed designs aimed at improving thermal management and adsorption efficiency. This demonstrates a key advantage of the present 3D CFD model in evaluating realistic and optimized adsorption system layouts for practical innovative future gas separation and $CO2$ capture applications.

Figure \ref{fig:complex_geometry} shows that in this new bed design configuration, the  reference bed design of figure \ref{fig:Mesh_Cyl}a is now replaced by seven smaller cylindrical tubes or channels. They are all vertically aligned and connected between the inlet and outlet, $28.2~mm$ each, as same inlet/outlet of the reference bed design shown in figure \ref{fig:Mesh_Cyl}a. The total internal volume is preserved to ensure the same quantity of adsorbent material as the reference case (figure \ref{fig:MeshCyl}), allowing thus for a direct comparison of performance metrics. The small seven cylinders, each has a diameter of 10.2 $mm$, are grouped within a diameter 35 $mm$ and a length 64 $mm$ of cylindrical column (see figure \ref{fig:complex_geometry}a).

A key design feature of this new geometry is the inter-cylinder spacing, which enhances thermal exchange with the surrounding environment while minimizing lateral thermal interference between adjacent cylinders. This layout promotes localized heat dissipation and reduces the risk of thermal accumulation often seen in monolithic designs. The computational mesh is carefully refined, ensuring that at least four cells span the interstitial gap between neighboring cylinders, allowing the CFD solver to accurately resolve thermal and mass transport phenomena at the interface regions. This new design geometry demonstrates the potential for geometrical optimization in adsorption bed design to improve process efficiency and local temperature regulation inside the bed.

\begin{figure}
    \centering
        \includegraphics[width=0.5\textwidth]{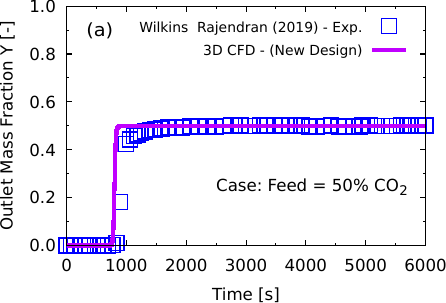}
        \includegraphics[width=0.5\textwidth]{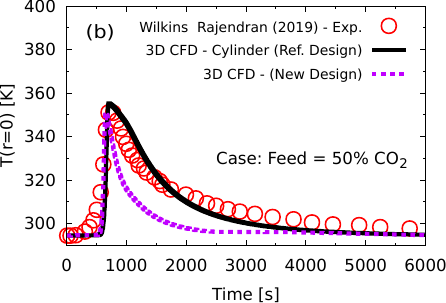}
    \caption{{The performance of the new bed design of figure \ref{fig:complex_geometry}, compared to the reference design of figure \ref{fig:MeshCyl} with its experimental data by Wilkins et al. 2019  \cite{wilkins2019measurement}. Case at 50\% CO\textsubscript{2} 50\% He gas mixture feed-in at $5.25 \cdot 10^{-6}~{m^3}/s$ flow rate and atmospheric pressure of $p=1.02$ bar. Temperature profiles correspond to the centerline local position (r=0) located at $z=52~mm$ away from the bed inlet. (a) Outlet CO\textsubscript{2} mass fraction (Y) as a function of time in seconds, and (b) local temperature T at $r=0$ and $z=52~mm$ inside the bed as function of time in seconds.}}
    \label{fig:mass_temp_validation_new}
\end{figure}

To further evaluate the robustness and adaptability of the 3D CFD framework, the validated simulation setup was extended to a more complex geometry composed of seven interconnected cylinders. This new configuration maintains the same total solid volume and adsorbent mass as the original cylindrical bed, thus preserving the adsorption capacity and breakthrough behavior. As seen in figure \ref{fig:mass_temp_validation_new}a, the outlet CO\textsubscript{2} mass fraction curve remains essentially unchanged, indicating that the new bed design does not compromise the saturation time or the amount of CO\textsubscript{2} captured.

However, the thermal response reveals a noteworthy improvement. While the peak temperature associated with the exothermic adsorption remains nearly identical (within 2 K) as shown in figure \ref{fig:mass_temp_validation_new}b, the post-adsorption cooling phase is significantly accelerated in the new bed design configuration. This enhanced thermal dissipation is attributed to the increased external surface area and spatial separation between individual cylinders, which reduces thermal coupling and promotes faster heat exchange with the surroundings.

This improvement in passive cooling implies that the bed can return to its initial thermal state more quickly, thereby reducing the required regeneration downtime. Consequently, the system can potentially support a higher number of adsorption-desorption cycles within a given operational period, effectively doubling the productivity in some use cases. These findings emphasize the critical role of geometry not only in adsorption efficiency but also in thermal management, an aspect that is often inaccessible to 1D or symmetry-constrained solvers.

\begin{figure}
    \centering
    \includegraphics[width=0.48\textwidth]{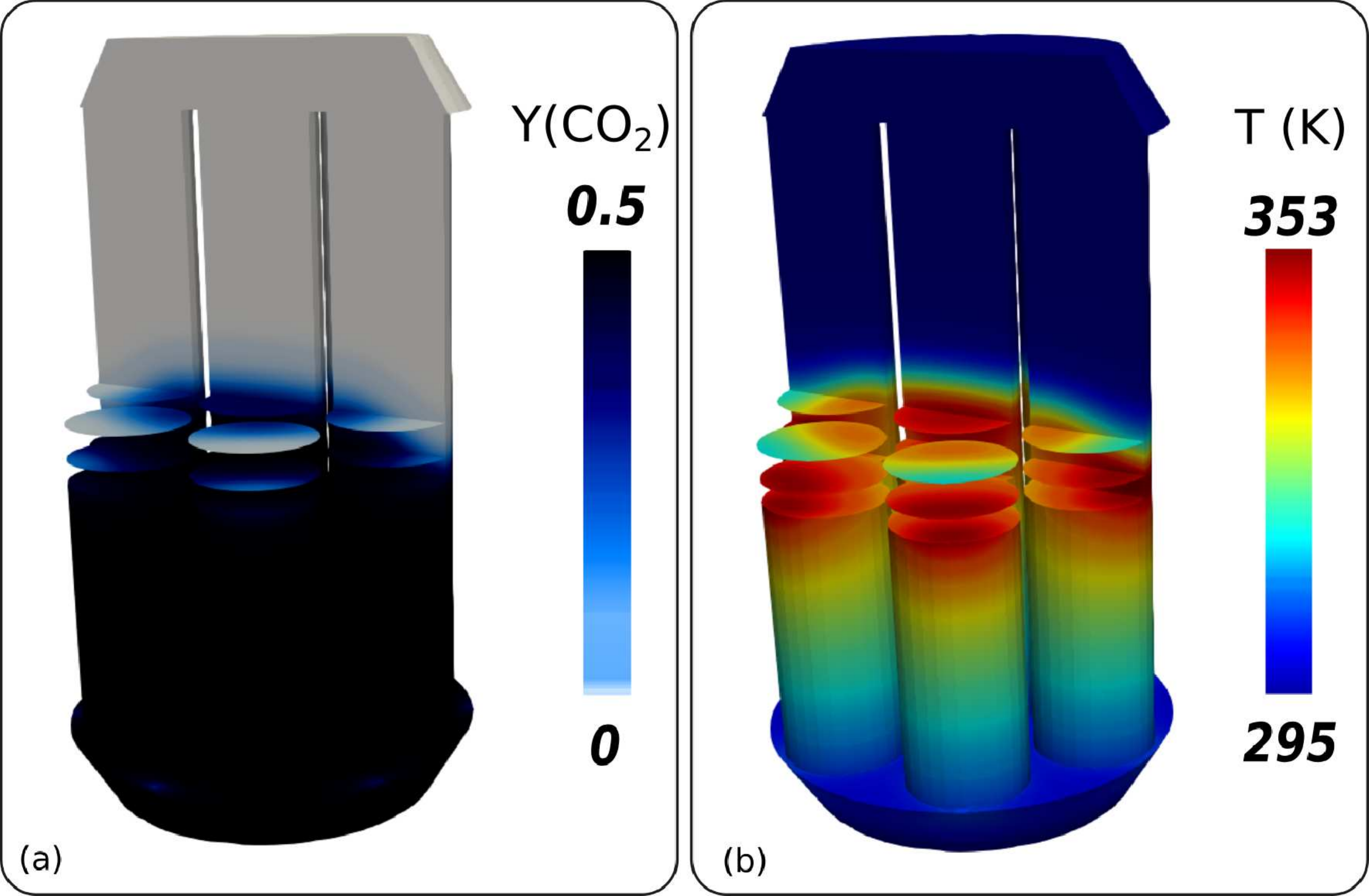}
    \caption{{3D CFD simulation results for the new bed design of seven inter-spaced tubes. CFD case under 50\% CO\textsubscript{2} 50\% He gas mixture feed-in and atmospheric pressure ($1.02$ bar) conditions. Results at time instance $t = 450~s$; (a): CO\textsubscript{2} mass fraction distribution and (b): temperature distribution.}}
    \label{fig:NewGeom3DResults}
\end{figure}

Figure~\ref{fig:NewGeom3DResults} presents the 3D simulation results for the new bed design made of seven identical vertical tubes that are inter-spaced between the bed's inlet and outlet. Each tube is individually packed with the adsorbent material. Despite the complex shape, the internal configuration was designed to preserve the same total adsorbent volume and material quantity as the reference monolithic cylindrical bed, ensuring a meaningful basis for comparison.

At $t = 450~s$, the CO\textsubscript{2} mass fraction field, figure \ref{fig:NewGeom3DResults}a shows a consistent and well-developed adsorption front progressing axially, uniformly distributed among the cylinders. This symmetry in concentration profiles confirms the validity of the inlet configuration and the physical homogeneity of the adsorption process across the entire bed.

In figure \ref{fig:NewGeom3DResults}b, the temperature distribution highlights the impact of the exothermic adsorption reaction. The localized temperature rise is clearly visible within the central zones of each sub-cylinder, while the outer walls exhibit cooler regions due to enhanced radial heat dissipation. This behavior illustrates one of the key advantages of the multi-cylinder design: increased surface area in contact with the ambient environment significantly accelerates heat removal.

Overall, the geometry enables an efficient decoupling of thermal zones, thereby reducing inter-cylinder thermal interaction and promoting faster cooling dynamics. The observed behavior confirms that this design maintains the adsorption performance of the original geometry, matching breakthrough times and adsorption capacity, while offering superior thermal management. This makes the system particularly attractive for high-frequency cyclic operation, as it allows quicker thermal regeneration between adsorption cycles.

\begin{figure*}
    \centering
    \includegraphics[width=0.9\textwidth]{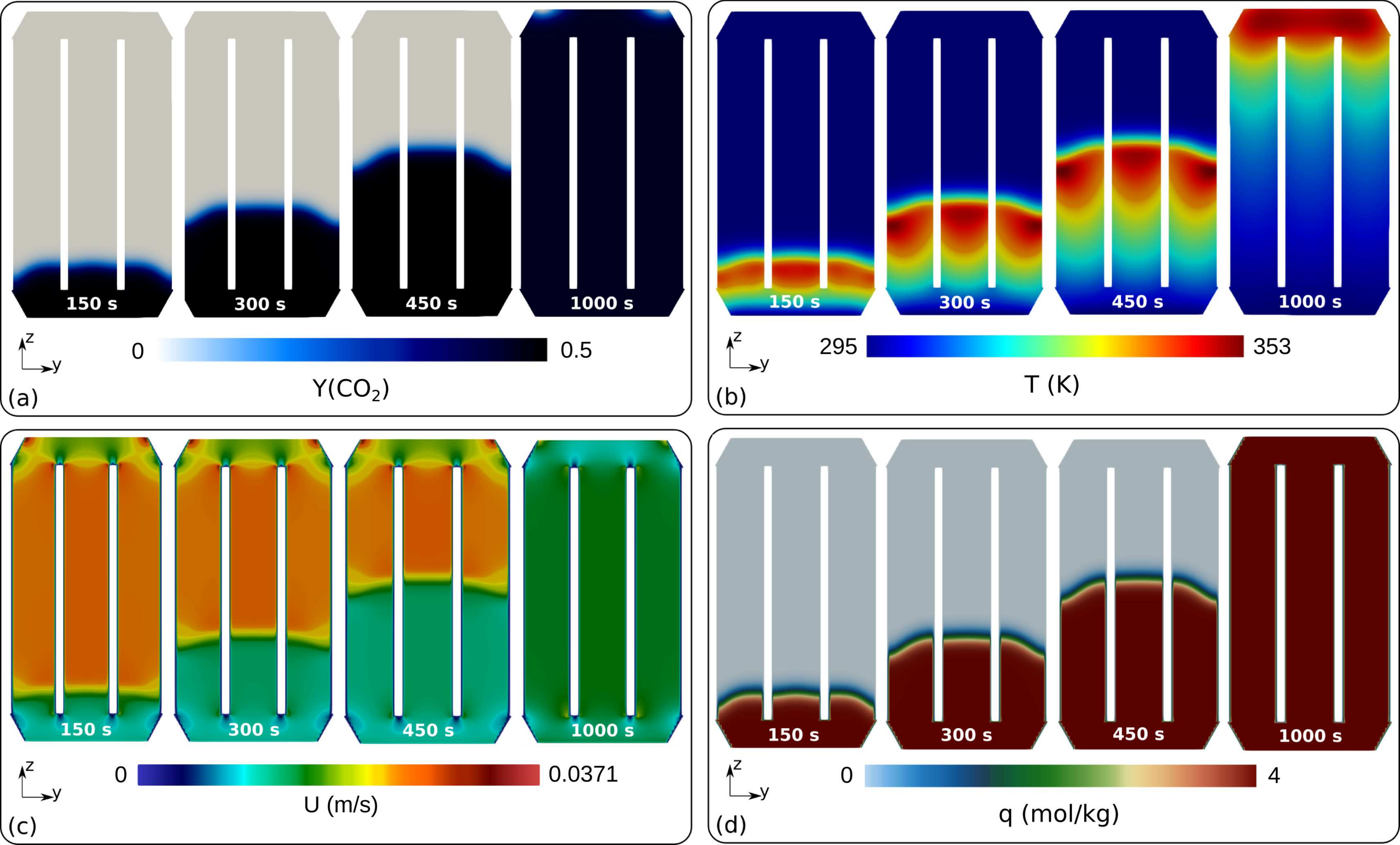}
    \caption{Space-time resolved present 3D CFD results for the new bed design geometry of figure \ref{fig:complex_geometry}. (a) CO\textsubscript{2} mass fraction (Y), (b) temperature $T$, (c) velocity magnitude $|\vec{U}|$, and (d) adsorbed CO\textsubscript{2} quantity $q$ (mol/kg). Results from left to right correspond to the time instances of 150 s, 300 s, 450 s and 1000 s.}
    \label{fig:comparison_time_fields_New_Geom}
\end{figure*}

Figure \ref{fig:comparison_time_fields_New_Geom} present cross-sectional slices perpendicular to the x-axis of new bed design (multi inter-spaced tubes of figure \ref{fig:complex_geometry}). They show the spatial distribution of key transport variables within the system at different time steps. This provides a detailed visualization of internal field variations across the interconnected channels at a given axial position. The scalar fields, such as velocity, CO\textsubscript{2} mass fraction $Y(CO_2)$, and the temperature can be observed inside the bed, highlighting the importance of full 3D CFD modeling. This helps to analyze new bed designs of complex geometry, and evaluate the effectiveness of local heat and mass transfer throughout the bed. Such 3D analysis is impossible using lower scale models like 1D models and/or 2D CFD models.

\begin{figure*}
    \centering
    \includegraphics[width=0.9\textwidth]{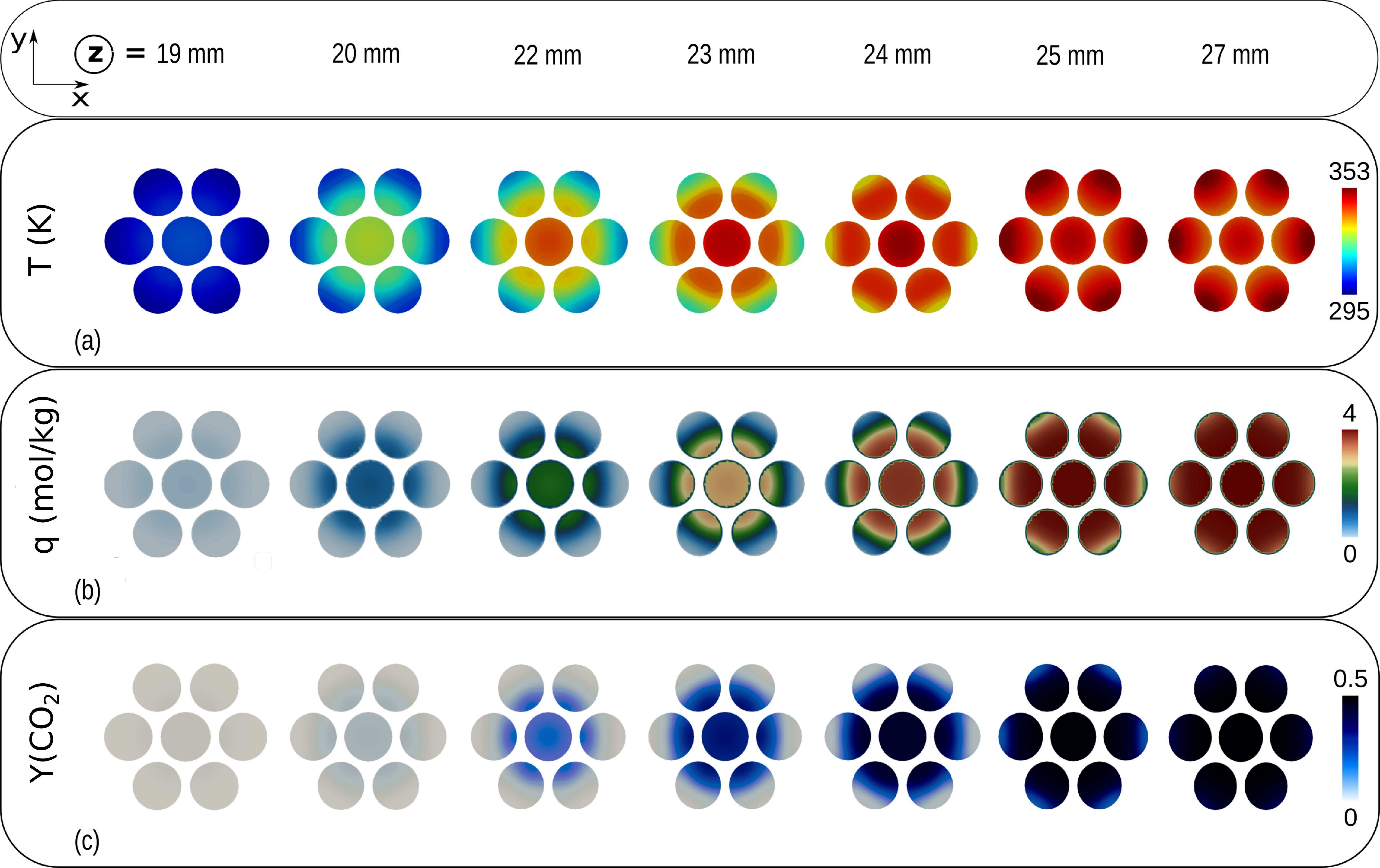}
    \caption{Space-time resolved present 3D CFD results for the new bed design geometry of figure \ref{fig:complex_geometry}. Showing at different z positions from inlet to outlet (xy-plane cross sections) of the new bed design. (a) the temperature $T$, (b) the adsorbed CO\textsubscript{2} quantity $q$ (mol/kg) and (c) the CO\textsubscript{2} mass fraction (Y). Results at the time instance of t=300 s.}
    \label{fig:front_propagation}
\end{figure*}

\begin{figure}
    \centering
    \includegraphics[width=0.48\textwidth]{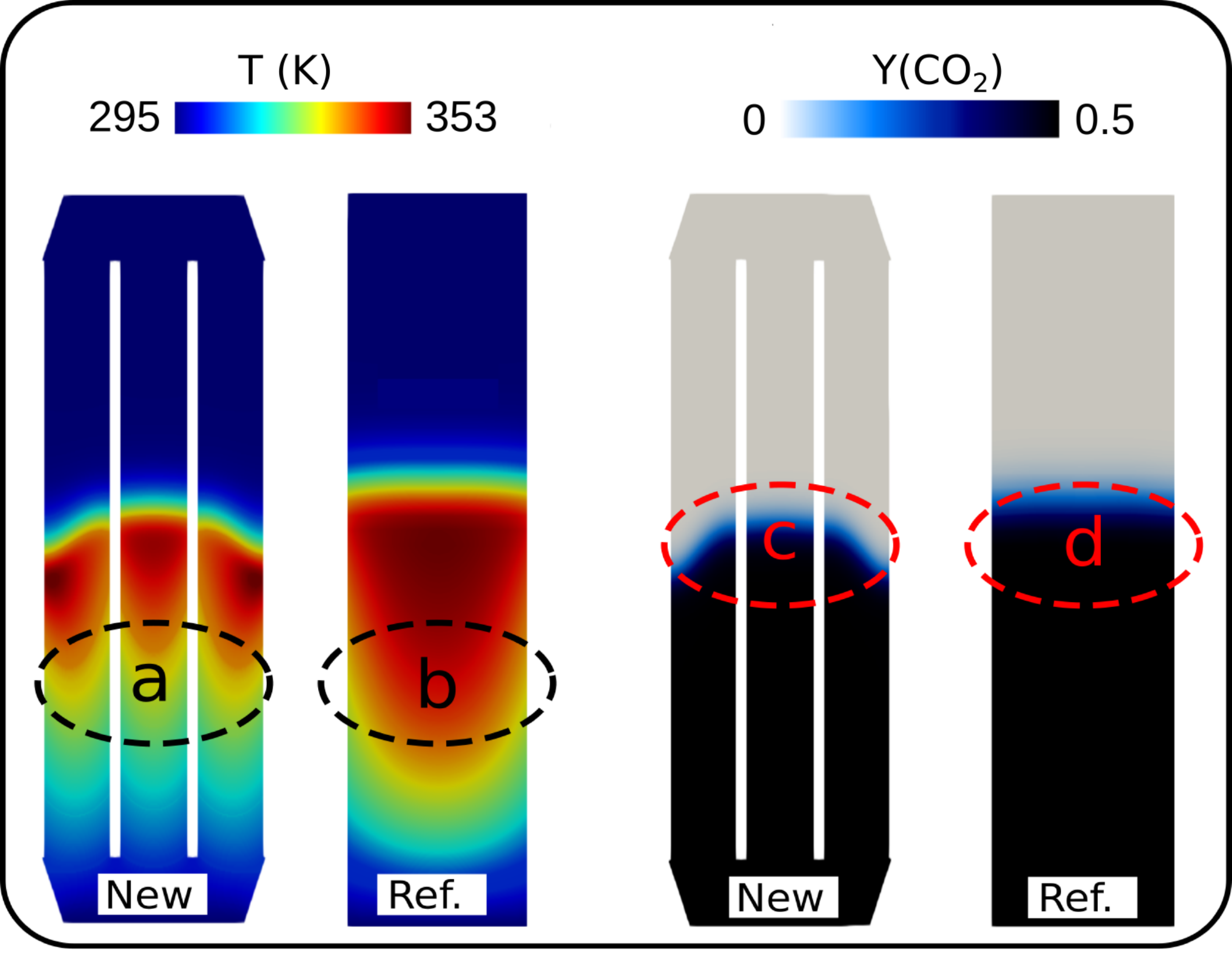}
    \caption{{Adsorbed CO$_2$ mass fraction and temperature fronts performance of: (a,c) the new bed design of figure \ref{fig:complex_geometry}; (b,d) the reference (Ref.) bed design of figure \ref{fig:Mesh_Cyl}. Temperature front at about the same time with: Dotted ellipse zone (a) shows a reduced local temperature in the bed compared to the zone (b). Adsorbed CO$_2$ mass fraction: Dotted ellipse zone (c) shows close adsorption front compared to the zone (d). }}
    \label{fig:comparison}
\end{figure}

Figure \ref{fig:front_propagation} present xy-plane cross-sections at different z positions inside the new bed design of figure \ref{fig:complex_geometry}. The temperature distribution ($T$), the adsorbed CO$_2$ quantity $q_0$, and the CO$_2$ mass fraction are shown at the time instance of t=300 s. These cross-sectional reveal the spatial evolution of thermal and mass transport phenomena within the new bed design, highlighting the complex interplay between fluid flow, heat transfer, and adsorption dynamics. Notably, temperature profiles indicate separation of the hot spots in adsorption front due to the multi-tube nature of the design. The adsorbed CO$_2$ quantity distribution elucidates the effectiveness of the adsorption kinematics across the bed length. These results, in addition to the CO$_2$ and $T$ local profiles in figure \ref{fig:mass_temp_validation_new}, provide valuable insight into the performance of the new bed design. Figures \ref{fig:mass_temp_validation_new}b and \ref{fig:comparison} clearly indicate the following: compared to a conventional bed made of single 3D cylinder (figure \ref{fig:Mesh_Cyl}), the present design of seven parallel cylinders (figure figure \ref{fig:complex_geometry}) significantly reduces the cooling duration of the adsorption period. This is extremely important in future PSA process to increase gas separation overall productivity, and thus enhance the overall PSA process efficiency.

\section{Conclusion and Perspectives}
\label{sec:conclusion}

This research presented the mathematical formulation, implementation, and validation of a new robust 3D CFD model for simulating gas adsorption phenomena in fixed-bed adsorbers, utilizing the open-source software platform \href{https://www.openfoam.com}{\textbf{OpenFOAM}}. The developed model couples the conservation equations for mass, momentum, and energy with adsorption kinetics, thus enabling both three-dimensional and temporally resolved representation of adsorption kinetics in fixed-bed adsorbers or reactors.

A central contribution of this work lies in taking into account the impact of pores adsorption occupation rate (PAOR) in the 3D CFD model. This is through a new volumetric source terms $\Gamma_Y$ in the gas species transport equation (\ref{equ:speciesConservation}) and $\Gamma_T$ in the energy conservation equation (\ref{equ:energyConservation}).

The 3D transient CFD model was validated using experimental data for three different  CO\textsubscript{2} feed-in concentrations (100\%, 50\%, and 15\%). The results showed excellent agreement in both the outlet gas composition and axial temperature evolution with time. This confirms the reliability of the present new 3D CFD model in reproducing both mass transfer and thermal behaviors under varying operating conditions of CO$_2$ capture by zeolite beads/pellets porous material.

Moreover, the new CFD model was applied to investigate the performance of a newly proposed fixed-bed adsorber design (a multi inter-spaced tubes configuration, i.e. see figure \ref{fig:complex_geometry}), maintaining equivalent adsorbent volume as in the reference single-tube cylindrical bed design (figure \ref{fig:Mesh_Cyl}), but with increased external surface area to enhance thermal dissipation. The 3D CFD simulation results revealed that while the breakthrough time and adsorption maximum local peak temperature remained effectively unchanged, the cooling duration of the adsorption cycle was substantially reduced. This reduces the duration of cyclic periods of CO$_2$ capture; i.e. within PSA and/or TSA processes. This confirms that 3D geometric optimization of fixed-bed reactors is very dependent on 3D CFD modeling in order to propose new generations of enhanced bed adsorbers; i.e. essential in toxic gas components removal technologies.

Despite the high fidelity of the current macroscopic CFD model, certain limitations persist. Chief among them is the assumption of a homogeneous porous adsorbing medium where the local distribution of adsorption quantity inside the particles is assumed to be uniform. As perspectives, further CFD models can be developed but at a lower scale, such as the scale of the adsorbing pellets/beads. In a coming future research, we will be for example focusing on the development of a meso-scale modeling framework in CFD, in which adsorption at a the scale of each individual particle is explicitly resolved within the 3D simulation domain. This will enable the assignment of non-uniform local physical properties to each adsorbing particle, thus allowing one to investigate anisotropy, particles shape and their orientation inside the bed. This will allow for a more detailed and mechanistic treatment of adsorption physics in fixed-bed adsorbers. Coupling such meso-scale particle-resolved modeling with the gas dynamics inside the bed will take the physics-background of the CFD model to a new scale/level.


\section*{Declaration of interest}
All the authors have no declaration of interest to address.

\section*{Data availability}
The data supporting this study's findings are available on request from the authors.

\section*{Acknowledgments}

The authors would like to thank the editorial board of Physics of Fluids,the Editor-in-Chief and the Journal's Managers and the AIP staff for their assistance during the review process of the present research. 

\section*{Authors Contribution} 
\textbf{M.N.} Conceptualization, Literature Review, Methodology, Software, Visualization, Original Draft; 
\noindent \textbf{M.S.} Conceptualization, Literature Review, Methodology, Supervision, Review Draft; 
\noindent \textbf{T.D.} Conceptualization, Literature Review, Methodology, Software, Code Development, Supervision, Review Draft, Project Administration;

\section*{Funding}
The authors thank the University and the Region of Rouen Normandy for funding this research under Project OMAC (\textbf{O}ptimization of \textbf{M}ulticomponent \textbf{A}dsorption \textbf{C}olumns for enhanced carbon capture and hydrogen production).

\section*{References}
\bibliography{references.bib}

@article{Songolzadeh2014,
author = {Songolzadeh, Mohammad and Soleimani, Mansooreh and Takht Ravanchi, Maryam and Songolzadeh, Reza},
title = {Carbon Dioxide Separation from Flue Gases: A Technological Review Emphasizing Reduction in Greenhouse Gas Emissions},
journal = {The Scientific World Journal},
volume = {2014},
number = {1},
pages = {828131},
doi = {https://doi.org/10.1155/2014/828131},
year = {2014}
}

@article{RIBOLDI20172390,
title = {Overview on Pressure Swing Adsorption (PSA) as {CO}2 Capture Technology: State-of-the-Art, Limits and Potentials},
journal = {Energy Procedia},
volume = {114},
pages = {2390-2400},
year = {2017},
note = {13th International Conference on Greenhouse Gas Control Technologies, GHGT-13, 14-18 November 2016, Lausanne, Switzerland},
issn = {1876-6102},
doi = {https://doi.org/10.1016/j.egypro.2017.03.1385},
author = {Luca Riboldi and Olav Bolland},
}

@article{SIQUEIRA20172182,
title = {Carbon Dioxide Capture by Pressure Swing Adsorption},
journal = {Energy Procedia},
volume = {114},
pages = {2182-2192},
year = {2017},
note = {13th International Conference on Greenhouse Gas Control Technologies, GHGT-13, 14-18 November 2016, Lausanne, Switzerland},
issn = {1876-6102},
doi = {https://doi.org/10.1016/j.egypro.2017.03.1355},
author = {Rafael M. Siqueira and Geovane R. Freitas and Hugo R. Peixoto and Jailton F. do Nascimento and Ana Paula S. Musse and Antonio E.B. Torres and Diana C.S. Azevedo and Moises Bastos-Neto},
}

@article{White2016,
title = {Development of a Pressure Swing Adsorption (PSA) Cycle for CO2 Capture From Flue Gas Using a 4-Bed {PSA} Apparatus},
year = {2016},
author = {Joshua White},
volume = {},
pages = {},
journal = {PhD Thesis}
}

@article{da1999general,
  title={A general package for the simulation of cyclic adsorption processes},
  author={Da Silva, Francisco A and Silva, Jos{\'e} A and Rodrigues, Al{\'\i}rio E},
  journal={Adsorption},
  volume={5},
  number={3},
  pages={229--244},
  year={1999},
  publisher={Springer}
}

@article{Zhang2025,
    author = {Zhang, Zhen and Liu, Gaofeng and Liu, Huan and Wang, Xiaoming and Lin, Jia and Barakos, George and Chang, Ping},
    title = {Fractal dynamics model of gas adsorption in porous media},
    journal = {Physics of Fluids},
    volume = {37},
    number = {1},
    pages = {016623},
    year = {2025},
    month = {01},
    issn = {1070-6631},
    doi = {10.1063/5.0250888},
}

@article{Zhou2025,
    author = {Zhou, Yu and Xiao, Aolei and Deng, Guanzheng and Li, Bohao and Lu, Xinlong and Li, Xiaoping and Wang, Jiale and Jing, Dengwei},
    title = {A novel fractal Langmuir and machine learning framework for precise prediction of CH4/CO2 adsorption in heterogeneous shale: Implications for CO2 sequestration},
    journal = {Physics of Fluids},
    volume = {37},
    number = {8},
    pages = {086639},
    year = {2025},
    month = {08},
    issn = {1070-6631},
    doi = {10.1063/5.0278272},
}

@article{carmo2020recovery,
  title={Recovery of vinyl chloride from by-streams of polyvinyl chloride production by TPSA in a multitubular adsorber},
  author={Carmo, Paulo and Ribeiro, Ana M and Rodrigues, Al{\'\i}rio E and Ferreira, Alexandre},
  journal={AIChE Journal},
  volume={66},
  number={5},
  pages={e16899},
  year={2020},
  publisher={Wiley Online Library}
}

@Article{LIU:2024,
title = {Research Progress of Pressure Swing Adsorption {CO}2 Capture Technology and Case Analysis of Its Application in Petrochemical Industry},
journal = {Southern Energy Construction},
volume = {11},
number = {5},
pages = {37-49},
year = {2024},
issn = {2095-8676},
doi = {10.16516/j.ceec.2024.5.04},	
author = {LIU Qiang and XIAO Jin and YU Hang and LUO Haizhong and HE Qingyang and LIN Haizhou and XUE Rong}
}

@article{kashefi2021point,
  title={Point-cloud deep learning of porous media for permeability prediction},
  author={Kashefi, Ali and Mukerji, Tapan},
  journal={Physics of Fluids},
  volume={33},
  number={9},
  year={2021},
  publisher={AIP Publishing}
}

@article{verbruggen2016cfd,
  title={{CFD} modeling of transient adsorption/desorption behavior in a gas phase photocatalytic fiber reactor},
  author={Verbruggen, Sammy W and Keulemans, Maarten and van Walsem, Jeroen and Tytgat, Tom and Lenaerts, Silvia and Denys, Siegfried},
  journal={Chemical Engineering Journal},
  volume={292},
  pages={42--50},
  year={2016},
  doi={10.1016/j.cej.2016.02.014},
  publisher={Elsevier}
}

@book{Yang1984,
  title     = {Gas Separation by Adsorption Processes},
  author    = {Yang, Ralph T.},
  year      = {1984},
  publisher = {Butterworth-Heinemann},
  address   = {Stoneham, MA, USA},
  isbn      = {978-0-408-11061-4}
}

@book{Ruthven1994,
  title     = {Pressure Swing Adsorption},
  author    = {Ruthven, Douglas M. and Farooq, S. and Knaebel, K. S.},
  year      = {1994},
  publisher = {VCH Publishers},
  address   = {New York, NY, USA},
  isbn      = {978-1-56081-562-4}
}

@book{welty2014fundamentals,
  title={Fundamentals of momentum, heat, and mass transfer},
  author={Welty, James and Rorrer, Gregory L and Foster, David G},
  year={2014},
  publisher={John Wiley \& Sons},
  issn = {9781118947463},
  address = {London}
}

@article{ergun1952,
  author    = {Ergun, Sabri},
  title     = {Fluid Flow Through Packed Columns},
  journal   = {Chemical Engineering Progress},
  volume    = {48},
  pages     = {89--94},
  year      = {1952}
}

@article{wilkins2019measurement,
  title={Measurement of competitive CO2 and N2 adsorption on Zeolite 13X for post-combustion CO2 capture},
  author={Wilkins, Nicholas Stiles and Rajendran, Arvind},
  journal={Adsorption},
  volume={25},
  number={2},
  pages={115--133},
  year={2019},
  doi={10.1007/s10450-018-00004-2},
  publisher={Springer}
}

@article{Naidu2021,
title = {Linear driving force analysis of adsorption dynamics in stratified fixed-bed adsorbers},
journal = {Separation and Purification Technology},
volume = {257},
pages = {117955},
year = {2021},
issn = {1383-5866},
doi = {https://doi.org/10.1016/j.seppur.2020.117955},
author = {Haripriya Naidu and Alexander P. Mathews},
}

@article{REZAEI2009,
title = {Optimum structured adsorbents for gas separation processes},
journal = {Chemical Engineering Science},
volume = {64},
number = {24},
pages = {5182-5191},
year = {2009},
issn = {0009-2509},
doi = {https://doi.org/10.1016/j.ces.2009.08.029},
author = {Fateme Rezaei and Paul Webley},
}

@article{glueckauf1955theory,
  title={Theory of chromatography. Part 10.—Formul{\ae} for diffusion into spheres and their application to chromatography},
  author={Glueckauf, Eugen},
  journal={Transactions of the Faraday Society},
  volume={51},
  pages={1540--1551},
  year={1955},
  publisher={Royal Society of Chemistry}
}

@book{keller2005gas,
  title={Gas adsorption equilibria: experimental methods and adsorptive isotherms},
  author={Keller, J{\"u}rgen U and Staudt, Reiner},
  year={2005},
  publisher={Springer Science \& Business Media},
  issn = {-- },
  address = {--}
}

@article{Celik2008,
    author = {Ismail B. Celik},
    title = {Procedure for Estimation and Reporting of Uncertainty Due to Discretization in CFD Applications},
    journal = {Journal of Fluids Engineering},
    volume = {130},
    number = {7},
    pages = {078001},
    year = {2008},
    month = {07},
}

@article{Aoming2025,
    author = {Aoming, Li and Peng, Cui and Xu, Wang and Adrian, Fisher and Lanyu, Li},
    title = {The artificial intelligence-catalyst pipeline: accelerating catalyst innovation from laboratory to industry},
    journal = {Front. Chem. Sci. Eng. },
    volume = {19},
    number = {55},
    pages = { },
    year = {2025},
    doi = {10.1007/s11705-025-2560-3},
}

@article{roache1994perspective,
    author = {Roache, P. J.},
    title = {Perspective: A Method for Uniform Reporting of Grid Refinement Studies},
    journal = {Journal of Fluids Engineering},
    volume = {116},
    number = {3},
    pages = {405-413},
    year = {1994},
    month = {09},
    issn = {0098-2202},
    doi = {10.1115/1.2910291},
}

@article{ramos2024cfd,
  title={CFD-based model of adsorption columns: Validation},
  author={Ramos, Henry Steven Fabian and Baliga, Chinmay and Rajendran, Arvind and Nikrityuk, Petr A},
  journal={Chemical Engineering Science},
  volume={285},
  pages={119606},
  year={2024},
  doi={10.1016/j.ces.2023.119606},
  publisher={Elsevier}
}

@article{gautier2018pressure,
title = {Pressure-swing-adsorption of gaseous mixture in isotropic porous medium: Transient 3{D} modeling and validation},
journal = {Chemical Engineering Journal},
volume = {348},
pages = {1049-1062},
year = {2018},
issn = {1385-8947},
doi = {https://doi.org/10.1016/j.cej.2017.05.145},
author = {R. Gautier and T. Dbouk and M.A. Campesi and L. Hamon and J.-L. Harion and P. Pré},
}

@article{GAUTIER2018314,
title = {Pressure-swing-adsorption of gaseous mixture in isotropic porous medium: Numerical sensitivity analysis in CFD},
journal = {Chemical Engineering Research and Design},
volume = {129},
pages = {314-326},
year = {2018},
issn = {0263-8762},
doi = {https://doi.org/10.1016/j.cherd.2017.11.007},
author = {R. Gautier and T. Dbouk and J.-L. Harion and L. Hamon and P. Pré},
}

@article{kasai2023cfd,
  title={CFD Modeling of Adsorption Rate Prediction for Granular Activated Carbon Packed Bed},
  author={Kasai, Yuma and Jinbo, Yoshinori and Kamikawa, Hideya and Sanada, Toshiyuki},
  journal={Journal of Chemical Engineering of Japan},
  volume={56},
  number={1},
  pages={2172980},
  year={2023},
  publisher={Taylor \& Francis}
}

@article{HWANG1995,
title = {Fixed-bed adsorption for bulk component system. Non-equilibrium, non-isothermal and non-adiabatic model},
journal = {Chemical Engineering Science},
volume = {50},
number = {5},
pages = {813-825},
year = {1995},
issn = {0009-2509},
doi = {https://doi.org/10.1016/0009-2509(94)00433-R},
author = {Kye Soon Hwang and Jae Ho Jun and Won Kook Lee},
}

@book{doble1984perry,
  title={Perry’s chemical engineers’ handbook},
  author={Doble, Mukesh},
  year={1984},
  publisher={McGraw-Hill: New York, NY, USA},
  issn = {-- },
  address = {New York}
}

@article{haghpanah2013multiobjective,
  title={Multiobjective optimization of a four-step adsorption process for postcombustion CO2 capture via finite volume simulation},
  author={Haghpanah, Reza and Majumder, Aniruddha and Nilam, Ricky and Rajendran, Arvind and Farooq, Shamsuzzaman and Karimi, Iftekhar A and Amanullah, Mohammad},
  journal={Industrial \& Engineering Chemistry Research},
  volume={52},
  number={11},
  pages={4249--4265},
  year={2013},
  publisher={ACS Publications}
}

@article{Song2025,
  title={Granulation mechanism and CO2 capture performance of alkaline metal salt-promoted MgO sorbents},
  author={Jinbo, Song and Jieying, Jing and Jinpeng, Zhang and Yufeng, Xu and Wen-Ying, Li},
  journal={Front. Chem. Sci. Eng.},
  volume={19},
  number={113},
  pages={},
  year={2025},
  doi = {10.1007/s11705-025-2576-8}
}

@article{vyas1995estimation,
  title={Estimation of Temperature-Dependent Thermal Conductivity of a Packed Bed of 13X Molecular Sieves},
  author={Vyas, Raj K and Kumar, Surendra},
  journal={Industrial \& engineering chemistry research},
  volume={34},
  number={11},
  pages={4058--4062},
  year={1995},
  doi={10.1021/ie00038a047},
  publisher={ACS Publications}
}

@article{ben2017multicomponent,
  title={Multicomponent and multi-dimensional modeling and simulation of adsorption-based carbon dioxide separation},
  author={Ben-Mansour, Rached and Basha, M and Qasem, Naef AA},
  journal={Computers \& Chemical Engineering},
  volume={99},
  pages={255--270},
  doi={10.1016/j.compchemeng.2017.01.040},
  year={2017},
  publisher={Elsevier}
}

@article{chen2017co2,
  title={{CO2} separation from offshore natural gas in quiescent and flowing states using 13X zeolite},
  author={Chen, SJ and Tao, ZC and Fu, Yue and Zhu, Mengmeng and Li, WL and Li, XD},
  journal={Applied Energy},
  volume={205},
  pages={1435--1446},
  year={2017},
  doi={10.1016/j.apenergy.2017.09.084},
  publisher={Elsevier}
}

@article{Hauchhum_Mahanta_2014, title={Carbon dioxide adsorption on zeolites and activated carbon by pressure swing adsorption in a fixed bed}, volume={5}, number={4 (December 2014)}, journal={International Journal of Energy and Environmental Engineering}, author={Hauchhum, Lalhmingsanga and Mahanta, Pinakeswar}, year={2014}, month={Aug.}, doi={10.1007/s40095-014-0131-3}}

@article{son2018,
  author = {Son, Y. and others},
  title = {Measurement and Prediction of the Heat of Adsorption and Equilibrium Concentration of {CO2} on Zeolite 13X},
  journal = {Energy \& Fuels},
  year = {2018},
  volume = {32},
  pages = {1663-1674}
}

@article{meng2026reactive,
  title={Reactive dispersion process in packed tube flow with wall adsorption and desorption},
  author={Meng, Linyue and Mei, Xiaoguang and Jiang, Weiquan},
  journal={Physics of Fluids},
  volume={38},
  number={1},
  year={2026},
  publisher={AIP Publishing}
}

@article{zou2025fractal,
  title={Fractal model of effective thermal conductivity in unsaturated porous media with rough-surfaced asymmetric tortuous tree-like bifurcating networks},
  author={Zou, Haoqian and Yang, Shanshan and Wang, Huili and Zheng, Qian},
  journal={Physics of Fluids},
  volume={37},
  number={7},
  year={2025},
  publisher={AIP Publishing}
}

\end{document}